\documentclass[aps,prl,twocolumn]{revtex4-2}
\usepackage{amssymb, amsmath, mathtools, amssymb}

\usepackage{graphicx}

\begin{document}

\title{Nonlinear Response of Diffusive Superconductors to $ac$-electromagnetic Fields}
\author{Pascal Derendorf, Anatoly F. Volkov, and Ilya M. Eremin}
\affiliation{Institut fur Theoretische Physik III, Ruhr-Universit\"at Bochum, 44801 Bochum, Germany}

\begin{abstract}
Motivated by the recent experimental progress in studying conventional and unconventional superconductors in a pump-probe setup, we perform a comprehensive theoretical analysis of the nonlinear response of a diffusive BCS conventional superconductor to the action of an alternating electromagnetic field using a generalized Usadel equation. We analyze the response up to the second order of the perturbation in the amplitude of
the vector potential $\vec{A}$, the superconducting order parameter $\Delta$ and in the third order for the current $\vec{j}$. On the basis of this approach, we derive general expressions for the retarded (advanced) Green’s functions, as well
as the Keldysh function for an arbitrary number of harmonics of the incident field. Most importantly, we analyze the set of physical observables in a non-equilibrium superconductor, such as frequency and the temperature dependencies of the zero harmonic $\delta( \Delta)_0$ (Eliashberg effect), the second harmonic $\delta( \Delta)_{2\Omega}$, as well as the third harmonic for the electric current
$j(3\Omega)$ under the action of a monochromatic
irradiation. For the same set of parameters, we also analyze the behavior of the reflectivity and the down-conversion intensity of a thin superconducting film, discussed recently in the context of parametric amplification of superconductivity. We derive these quantities microscopically and show the connection of the down-conversion intensity to the third harmonic generation currents induced by the amplitude mode and the direct action of the electric field on the charge carriers.
\end{abstract}

\maketitle
\date{\today }

\section{ Introduction}

The study of non-stationary properties of superconductors started soon after
the formulation of the BCS theory of superconductivity \cite{BCStheory57} with the
calculation of the ac conductance of superconductors \cite%
{MattisBardeen58,AbrikosovGorkov59}. Later, Gor'kov and Eliashberg \cite%
{GorkovEliashberg68} investigated nonlinear ac properties on the basis of
generalized Ginzburg-Landau equations. The ac conductance in a
current-carrying superconductors was studied in Refs. \cite%
{GarfunkelPRL66,GarfinkelPRB73,OvchinnikovIsaakyan78} and more recently in Refs.\cite{Moor2017,Crowley22,Poniatowski2022,Jujo2023} using more microscopic approach. 

For obvious reasons one of the most interesting problems is to study
the spectrum of collective modes of the phase $\varphi $ and amplitude mode $%
\delta \Delta $ of the superconducting order parameter $\Delta $. In the
absence of a microscopic theory, these modes were discussed by
Anderson \cite{AndersonPR58a,AndersonPR58a} and independently by Bogoljubov\cite{Bogoljubov58}.
In particular, Anderson came to the conclusion that phase variations in
space and time (phase mode) $\varphi (t,x)$ lead to perturbations of the
electric charge and, therefore, cannot exist in metals until the frequency $%
\omega $ exceeds the plasma frequency $\omega _{pl}$, which is much higher
than $\Delta $. Remarkably, this idea of Anderson motivated Higgs in his
prediction of a new type of boson (Higgs boson) in particle physics \cite%
{HiggsPRL64,Volovik13}. 

Contrary to that, the amplitude mode (AM) is electrically neutral and can be excited in the absence of an additional interaction, for example, the Coulomb interaction. The AM in BCS superconductors corresponds to perturbations of the quasiparticle distribution function $n(\xi _{p})$ symmetric with respect to a variable $\xi _{p}$, here $
\xi _{p}=v_{F}(p-p_{F})$. The phase mode is caused by a
branch-imbalance, {\it i.e.} by a deviation from zero of a part of the function
$n(\xi _{p})$ asymmetric in variable $\xi _{p}$.\cite{Tinkham72,Schmid75,Artemenko1979}
The amplitude mode is confusingly called the Higgs mode, although logically the Higgs mode would rather correspond to the plasma mode. This fact has also been highlighted in
several articles.\cite{ShimanoRev20, Silaev22}. 

Theoretically, the
evolution of $\delta \Delta (t)$ after a sudden perturbation (quench) was studied in
Ref.\,\cite{VolkovKogan74} in a way similar to the problem of the electric
field $E$ evolution in a collisionless plasma considered by Landau \cite{Landau46}.
Landau showed that an initial perturbation of the electric field $E$ in plasma oscillates in time with plasma frequency $\omega _{pl}$ and
attenuates \textit{exponentially}  even in the absence of collisions.  
The same problem in a BCS superconductor leads to a different result. In the absence of inelastic
collisions, homogeneous perturbations of the order parameter $\delta \Delta
(t)$ in superconductors oscillate with a frequency of $2\Delta $, and also
decay in time as follows
\begin{equation}
\delta \Delta (t)\sim \delta \Delta _{exc}\frac{\cos (2\Delta t+\theta _{0})%
}{\sqrt{2\Delta t}}+\delta \Delta (\infty )  \label{In2}
\end{equation}%
where $\delta \Delta (\infty )/\Delta _{0}\neq \delta \Delta (0)$, i.\,e.\,$%
\delta \Delta (t)$ approaches not to initial value $\delta \Delta (0)$ but some steady (smaller) value.
Here, the order parameter $\Delta $ are expressed as an integral over momenta $p$ or
energies $\epsilon $ of the distribution function (the condensate wavefunction $f(\epsilon )$). The decay rate in Eq.\,(\ref{In2}) stems from the branch point singular points in the integral from the function $f(\epsilon )$, while the exponential attenuation of $E$ in plasma is due to the pole of the function $n(\xi _{p})$. The temporal behavior of $\delta \Delta (t)$ stemming from the branch point at $\epsilon =2\Delta $ was discussed by different methods for various systems not only in Refs.\cite{VolkovKogan74,Kulik81,Omelyanchuk04,Barankov2004,Papenkort2007,Akbari2013,Krull2014,Cui2019} for
BCS\ s-wave superconductors, but also in  for neutral fermionic superfluids\cite{Yuzbashyan2006a,Yuzbashyan2006b,Dzero2007,Kurkjian2019} or BCS-BEC (Bose-Einstein
condensate) crossover regime.\cite{Szymanska2005,Warner2005,Gurarie2009,Yuzbashyan2015,Scott2012,Seibold2020,Chubukov23}, as well as in proximitized structures\cite{Vadimov2019,Plastovets2023}.

The analogy between collisionless superconductors and plasma turned out to
be even deeper; both systems belong to the class of fully integrable
systems (in the absence of collisions). In these systems, information about
oscillations is conserved in electrons and is lost in order parameter $%
\delta \Delta (t)$ (in superconductors) or in electric field $E$ (in
plasma). Such a memory may lead to echo effects \cite{Kadomtsev1968}. The analogy with plasma was noted already in Refs.\cite{VolkovKogan74,Volkov75}. In particular, it was emphasized in Ref.\,\cite%
{Volkov75} that contrary to $\delta \Delta (t)$ the condensate wave function
$f(\epsilon ,t)$ experiences non-decaying oscillations. Further development
of plasma theory has shown that undamped plasma waves can also be excited in
plasma if the initial conditions are chosen appropriately \cite{Kadomtsev1968} and similar ideas have been further developed for superconductors
\cite{Barankov2004,Omelyanchuk04,Yuzbashyan2006a,Tsuji2015} and superfluids \cite{Szymanska2005,Warner2005,Dzero2007,Gurarie2009,Scott2012,Yuzbashyan2015,Seibold2020}. 

Experimental observation of the $\delta \Delta (t)$ dynamics is a rather
difficult experimental task since inelastic scattering times are short ($%
\tau _{in}\sim 10^{-9}-10^{-12}\sec $) and in the linear approximation the
amplitude mode does not directly couple with the electromagnetic field. Historically, the amplitude mode was first revealed in the
Raman scattering in a charge density wave superconductor NbSe$_2$ in 1981 \cite{KleinPRL80,LittlewoodVarma82} (see also recent works and references therein \cite{Cea14}).
In addition, early studies of superconducting properties under continuous irradiation with sub-gap energies using microwaves already showed surprising results that superconductivity could be observed above the equilibrium critical temperature, a phenomenon that is \cite{Wyatt66,Dayem67,Klapwijk77}. These observations were explained by Eliashberg, who showed that in the
vicinity of the critical temperature, the effect is mainly related to a non-equilibrium distribution of quasiparticles induced by the microwave field.\cite{Eliashberg70,Ivlev73} 

Furthermore, recent advances in the technical development of THz spectroscopy enabled the construction of THz-pump THz-probe spectroscopy in the conventional $s$-wave superconductor such as NbN and  Nb$_{1-x}$Ti$_x$N thin films, which reveal novel experimental observations. For example, transient gap enhancement, suggesting an additional Eliashberg effect present at temperatures close to T$_c$ was observed in NbN\cite{Beck13,Demsar2020}

Further studies concerned the observation of the collective modes followed. \cite{Matsunaga2013,Matsunaga2014} The idea of this experimental setup is that a single-cycle (pump) pulse with an energy on the order of $\Delta$ excites the amplitude mode in the superconducting condensate non-adiabatically. This leads to an amplitude oscillation of the order parameter $\Delta$ with the frequency ${\Omega_h = 2\Delta}$. These oscillations can be detected with a second, weaker pulse that comes with a delay $\delta t$ and probes these dynamics. The transmitted electric fields can be measured using electro-optic probing \cite{Matsunaga2014}. The probe pulse is applied with an orthogonal polarization to the pump pulse. Thus, the pump pulse can be filtered out from the probe pulse using a wire grid polarizer Ref.\,\cite{Matsunaga2012}.

While a non-adiabatic quench with a single-cycle pulse on the picosecond scale is an experimental challenge, one can consider to drive the system periodically with a multi-cycle pulse of frequency $\Omega$. Since the amplitude mode couples to light in the second order, it is driven at twice the driving frequency $2\Omega$. The driven oscillation can be then measured as an induced current. 
Since the electromagnetic vector potential oscillates with the driving frequency $\mathbf{A}(t)\sim e^{i\Omega t}$ and the amplitude mode oscillates with twice the driving frequency $\delta\Delta_{H}(t)\sim e^{2i\Omega t}$, one obtains a total current, which oscillates with three times the driving frequency $\mathbf{j}(t)\sim e^{3i\Omega t}$ \cite{Tsuji2015,Murotani2017}. This generation of a third-harmonic is a process, which only occurs inside the superconducting state at $T<T_c$ and should be resonantly enhanced once $2\Omega$ activates the amplitude mode at $2\Omega = \Omega_H = 2\Delta$.

In Ref.~\cite{Matsunaga2014} the first successful measurement of the third-harmonic generation in the conventional $s$-wave superconductor Nb$_{1-x}$Ti$_x$N was reported. However, soon after it was realized that the activation of the amplitude mode is not the only process measured in the current produced by the third-harmonic generation. Since the frequency of the amplitude mode is given at $\Omega_H=2\Delta$, it coincides with the onset of the particle-hole generation and with the energy needed to break the Cooper pairs. In particular, Ref.~\cite{Cea2016} has shown that in the clean BCS superconductor the contribution to the third harmonic generation current caused by a direct action of ac electric field (sometimes called charge density fluctuations, which is not completely adequate) is three orders of
magnitude larger than that from the amplitude mode. Subsequent studies analyzed how impurities, realistic electron-phonon coupling, and strong coupling features affect this ratio.\cite{Jujo18,Murotani2019,Silaev2019,Tsuji2020,Haenel2021,Seibold2021,Yang2022} 
Obviously, this shows that research on the detection of the collective modes and their behavior in a non-equilibrium in conventional and unconventional superconductors due to a driving field is still ongoing. Furthermore, several works were devoted to the study of possibilities to detect the amplitude mode from the nonlinear response of a superconductor irradiated terahertz electromagnetic fields. One of these methods is to measure the ac
conductance of a current-carrying superconductor $\sigma (\Omega )$ \cite{Moor2017,Poniatowski2022,Crowley22}. In this case, the electromagnetic field is coupled to the AM and the ac conductance $\sigma (\Omega )$ contains a term related to the excited  amplitude mode (Higgs mode) \cite{Moor2017}. 

If a superconductor is suddenly brought out of
equilibrium and the order parameter $\delta \Delta (t)$ oscillates, as follows from 
Eq.\,(\ref{In2}), the behavior of the system resembles the behavior of
a generator with a resonant frequency of $2\Delta $. It is, therefore, quite natural to
suppose that this system might be used as a parametric oscillator. The authors of Refs.\cite{Buzzi2021b,Buzzi2021PRX} on the basis of a phenomenological model, analyzed the possibility of parametric amplification
but microscopic analysis is still lacking. Most importantly, the idea of parametric amplification has been used among others to explain the highly interesting observation in high-T$_c$ cuprates\cite{Fausti2011,Nicoletti2014,Kaiser2014,Hu2013,Nicoletti2018,Liu2020}, K$_3$C$_{60}$\cite{Mitrano2016,Cantaluppi2017,Budden2021,Buzzi2021b,Rowe2023}, and certain organic superconductors\cite{Buzzi2020,Buzzi2021a}. In particular the intense far-infrared optical pulses have been shown to
create non-equilibrium coherent states with optical properties that are consistent with transient  photo-induced superconductivity phenomenon, yet the nature of this state is still debated.\cite{Katsumi2022,Dodge2023,zhang2023lightinduced}

\begin{figure} [htbp]
    \centering
    \includegraphics[width=0.41  \textwidth]{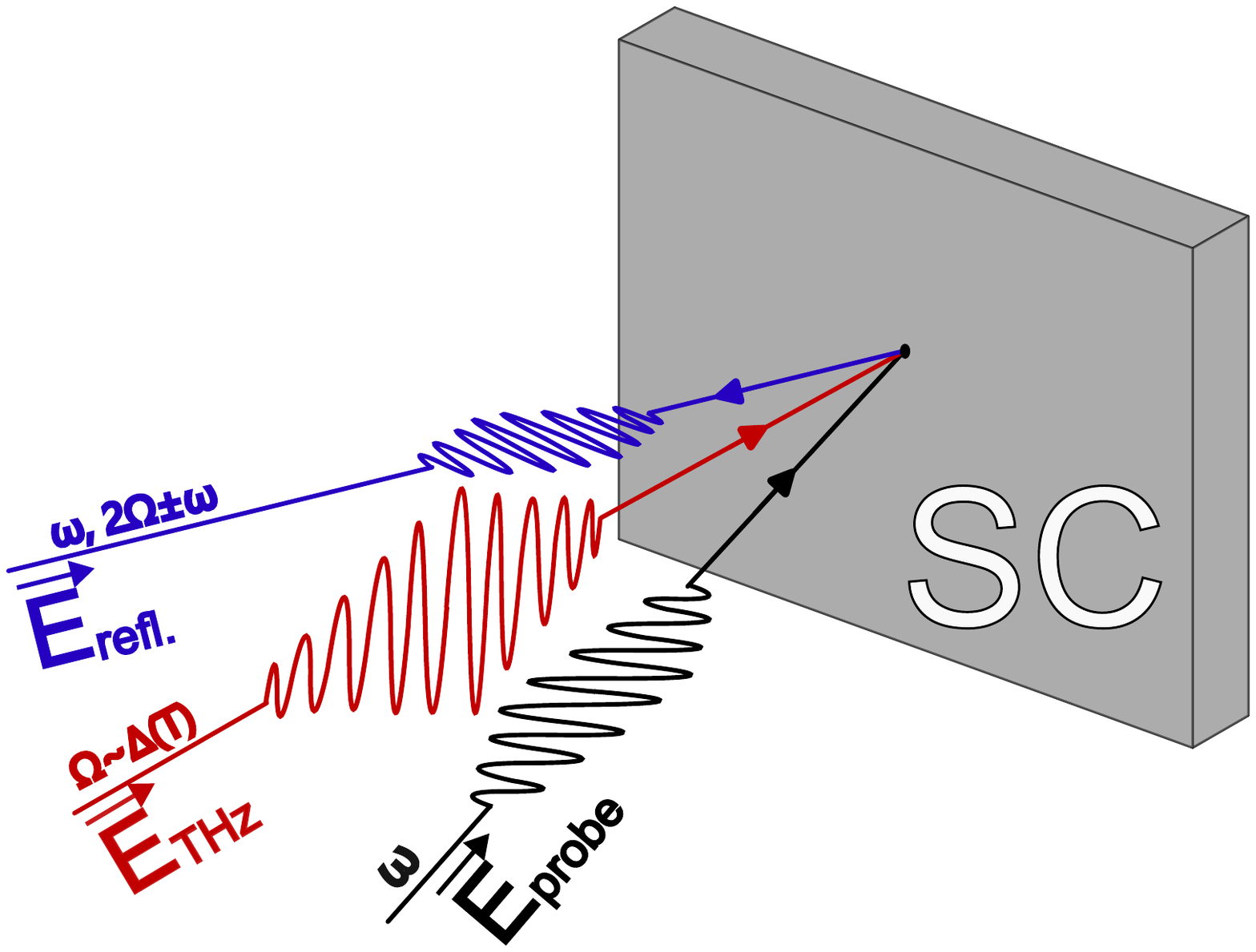}
    \caption{Schematic representation of the system under consideration. We consider the superconducting sample, which is continuously irradiated by the monochromatic THz pulse $\vec{E}_{\text{THZ}}(\Omega \sim \Delta(T))$. At the same time, the probe field is applied with orthogonal polarization $\vec{E}_{\text{probe}}(\omega)$. The reflected electromagnetic field, $\vec{E}_{\text{refl}}$ is measured either for the probe field frequency, $\omega$, or at $2\Omega\pm \omega$ to reveal the parametric effects.}
    \label{fig:Schematic}
\end{figure}

In this manuscript, we calculate within a microscopic quasiclassical description the response of
a superconductor $\delta \Delta (t)$ in the second order to the action of a
monochromatic irradiation or of many signals with different $\Omega $ following the experimental setup, presented in Fig.\,\ref{fig:Schematic}. In particular, we consider the superconducting sample, which is continuously irradiated by the monochromatic THz pulse $\vec{E}_{\text{THZ}}(\Omega \sim \Delta(T))$. In
the second order of the field intensity $|Q_{\Omega }|=eE_{\Omega }/\Omega $%
, we obtain general formulas for the quasiclassical Green's function $f(\epsilon ,t)$.  From these we find the variation $\delta \bar{\Delta}$ averaged in time
(Eliashberg effect), the variation $\delta \Delta _{2\Omega }$ at the
doubled frequency of a monochromatic ac field. We also calculate the third
harmonic of the current $I_{3\Omega }$ and the response $I_{p}$ on the action
of ac field with a frequency $\Omega _{p}$. Here we assume that the probe field is applied with orthogonal polarization $\vec{E}_{\text{probe}}(\omega<2\Delta(T))$. The reflected electromagnetic field, $\vec{E}_{\text{refl}}$ is measured either for the probe field frequency, $\omega$, or at $2\Omega\pm \omega$ to reveal the parametric gain (amplification) conditions
and a change in the ac conductance in the presence of a pump irradiation. Note, unlike the recent work\cite{kim2023tracing} we do not study a transient behavior of the superconductor irradiated by ac pulses. and assume that ac irradiation is continuous in time.

\section{General Equations}

To model a realistic experimental situation we consider diffusive superconductors ($\tau _{el}T_{c}\ll 1$). To describe non-stationary and
non-linear response of an $s$-wave conventional superconductor, we will use
a well-established theory for matrix quasiclassical Green's functions $\hat{g}$
\cite{Eilenberger68,UsadelPRL70,Eliashberg71,LO66,VZKlap93} (for review see also \cite{RammerSmithRMP86,ZaikinRev99,KopninBook01}). Equation for the
retarded $g^{R(A)}(t,t^{\prime })$ function has the form (we dropped the superindex $R(A)$)
\begin{equation}
i(\hat{\tau}_{3}\partial _{t}\hat{g}+\partial _{t^{\prime }}\hat{g}\hat{\tau}%
_{3})+[\hat{\Delta},\hat{g}]-iD\partial (\hat{g}(\partial \hat{g}))=0\,.
\label{1}
\end{equation}%
where $[\hat{\Delta},\hat{g}]=\hat{\Delta}(t)\hat{g}(t,t^{\prime })-\hat{g}
(t,t^{\prime })\hat{\Delta}(t^{\prime })$. In the presence of a
gauge-invariant ''momentum'' $\mathbf{Q}=\nabla \chi /2-(e/c)\mathbf{A}$,
related to the condensate velocity $v_{S}=Q/m$, Eq.\,(\ref{1}) acquires the
form
\begin{equation}
i(\hat{\tau}_{3}\partial _{t}\hat{g}+\partial _{t^{\prime }}\hat{g}\hat{\tau}%
_{3})+[\hat{\Delta},\hat{g}]=-iD[\hat{Q}\hat{g}\hat{Q},\hat{g}].  \label{2}
\end{equation}
\begin{widetext}
Here $\hat{Q}=Q\hat{\tau}_{3}$ and
\begin{eqnarray}
[Q\hat{\tau}_{3}\hat{g}Q\hat{\tau}_{3},
\hat{g}]=  
 \int dt_{1}\{\hat{Q}(t)\hat{g}(t,t_{1})\hat{Q}(t_{1})\hat{g}
(t_{1},t^{\prime })-\hat{g}(t,t_{1})\hat{Q}(t_{1})\hat{g}(t_{1},t^{\prime })
\hat{Q}(t')\} 
\end{eqnarray} 
\end{widetext}

In the considered spatially
uniform case, one can choose the phase $\chi $ equal to zero so that the
electric fields $\mathbf{E}$ is related to $\mathbf{Q}$ via $\mathbf{E}%
=-(1/c)\partial _{t}\mathbf{A}=\partial _{t}\mathbf{Q}/e$. We consider the
case when momentum $Q(t)$ consists of a sum of periodic functions
\begin{equation}
Q(t)=\sum_{\pm (\nu ,\mu )}[Q_{\nu }\exp (i\Omega _{\nu }t)+Q_{\mu }\exp
(i\Omega _{\mu }t)]  \label{2a}
\end{equation}
with arbitrary frequencies $\Omega _{\nu ,\mu }$. Because $Q(t)$ is a real
function, we require that $Q_{\nu ,\mu }=Q_{-\nu ,-\mu }$. In particular,
the frequency $\Omega _{\nu }$ can be equal to $\Omega _{\mu }$: $\Omega
_{\nu }=$ $\Omega _{\mu }=\Omega $ (monochromatic irradiation). For the
Fourier components $\hat{g}_{\epsilon ,\epsilon ^{\prime }}=\int dtdt^{\prime
}\hat{g}(t,t^{\prime })\exp (i\epsilon t-i\epsilon ^{\prime }t^{\prime })$,
Eq.\,(\ref{2}) becomes (see for example \cite{Artemenko1979,Moor2017,SilaevPRB19})
\begin{equation}
\epsilon \hat{\tau}_{3}\hat{g}-\hat{g}\hat{\tau}_{3}\epsilon ^{\prime }+[%
\hat{\Delta},\hat{g}]_{\epsilon,\epsilon'}=-iD[Q\hat{\tau}_{3}\hat{g}Q\hat{\tau}_{3},\hat{g}]_{\epsilon,\epsilon'}
\label{3}
\end{equation}
The right-hand side can be written as
\begin{widetext}
\begin{eqnarray}
iD[Q\hat{\tau}_{3}\hat{g}Q\hat{\tau}_{3},\hat{g}]_{\epsilon,\epsilon'}=
iD\sum_{\nu \mu }\{\hat{Q}
_{\nu }\hat{g}_{\epsilon +\Omega _{\nu }}\hat{Q}_{\mu }\hat{g}_{\epsilon
^{\prime }}-\hat{g}_{\epsilon }\hat{Q}_{\nu }\hat{g}_{\epsilon ^{\prime
}-\Omega _{\mu }}\hat{Q}_{\mu }\}2\pi\delta (\epsilon _{-}+\Omega _{\nu +\mu }).
\label{3a}
\end{eqnarray}%
\end{widetext}
with $\epsilon _{-}=\epsilon-\epsilon ^{\prime } $ and $\Omega _{\nu +\mu
}\equiv \Omega _{\nu }+\Omega _{\mu }$.

Consider first the unperturbed ground state. The retarded (advanced) Green's functions in the ground state have a
standard form
\begin{eqnarray}
\hat{g}_{0}^{R(A)} &=&g(\epsilon )\hat{\tau}_{3}+i\hat{\tau}_{2}f(\epsilon
)|^{R(A)}\text{,}  \label{4} \\
 g_{0}^{R(A)}(\epsilon ) &=&(\epsilon \pm i\gamma )/\zeta ^{R(A)}\text{, }%
f(\epsilon )=\Delta /\zeta ^{R(A)}\text{.}  \label{4a}
\end{eqnarray}%
where
\begin{equation}
\zeta ^{R(A)}(\epsilon )={\Big \{}%
\begin{array}{c}
\pm \text{sign}(\epsilon) \sqrt{(\epsilon \pm i\gamma )^{2}-\Delta ^{2}}\text{%
, }|\epsilon |>\Delta \\
i\sqrt{\Delta ^{2}-\epsilon ^{2}}\text{, }|\epsilon |<\Delta%
\end{array}%
\;.  \label{5}
\end{equation}%
Here $\gamma $ is a phenomenological damping coefficient introduced by
Dynes et al.\,\cite{DynesPRL78,Hlubina2018} which is assumed to be small ($\gamma \ll
\Delta $). One can see that $\zeta ^{A}(\epsilon )=\zeta ^{R}(-\epsilon )$
and $\zeta ^{A}(\epsilon )=-(\zeta ^{R}(\epsilon ))^{\ast }$.
At the next step we consider the corrections to $\Delta _{0}$ and to $\hat{g}_{0}^{R(A)}$ due to
ac perturbations.

\section{Action of the ac Fields}

In a nonequilibrium or non-stationary case, the system is described by a
matrix $\check{g}$ whose elements are the retarded (advanced), $\hat{g}%
^{R(A)}$, and the so-called Keldysh matrix function $\hat{g}^{K}$. The
supermatrix Green's function $\check{g}$ is defined as
\begin{equation}
\check{g}={\Big \{}%
\begin{array}{c}
\hat{g}^{R}\text{, }\hat{g}^{K} \\
0\text{ , }\hat{g}^{A}%
\end{array}%
{\Big \}}.  \label{5A}
\end{equation}%
where $\hat{g}^{R(A)}$ are the retarded\ (advanced) Green's functions and
and the Keldysh Green's function $\hat{g}^{K}$ \cite{Keldysh65}. The latter
matrix Green's function is expressed in terms of the matrix distribution
function $\hat{n}$ \cite%
{LO66,RammerSmithRMP86,VZKlap93,ZaikinRev99,KopninBook01}
\begin{equation}
\hat{g}^{K}=\hat{g}^{R}\hat{n}-\hat{n}\hat{g}^{A} , \label{5B}
\end{equation}%
and the matrix $\check{g}$ obeys the normalization condition
\begin{equation}
\check{g}\cdot \check{g}=\check{1}  \label{5C}
\end{equation}

The correction $\delta \check{g}(\epsilon ,\epsilon ^{\prime })=\check{g}%
(\epsilon ,\epsilon ^{\prime })-$ $\check{g}_{0}(\epsilon )2\pi \delta
(\epsilon -\epsilon ^{\prime })$ satisfies the linearised Eq.\,(\ref{3}) which
acquires the form 

\begin{equation}
\begin{split}
({\zeta} _{\epsilon }\check{g}_{\epsilon })\delta \check{g}-\delta \check{g}%
(\check{g}_{\epsilon ^{\prime }}{\zeta} _{\epsilon ^{\prime }})=\sum_{\nu ,\mu
}&\left[\check{R}_{\Delta }(\epsilon ,\epsilon ^{\prime })+\check{R}%
_{Q}(\epsilon ,\epsilon ^{\prime })\right] \\ 
&\times 2\pi\delta (\epsilon _{-}+\Omega
_{\nu +\mu }) \label{6}
\end{split}
\end{equation}
where the {\it rhs} contains only the Green's functions in the ground
state and the matrix elements of the $4\times 4$ matrices $\check{g}%
_{\epsilon }$ are $\hat{g}_{\epsilon }^{R(A)}=\hat{g}_{0}^{R(A}(\epsilon )$,
$\hat{g}_{0}^{K}=[\hat{g}_{0}^{R}(\epsilon )-\hat{g}_{0}^{A}(\epsilon
)]t_{\epsilon } $ with $t_{\epsilon }=\tanh (\epsilon \beta )$ and $\beta
=(2T)^{-1}$.
The supermatrix $(\zeta_\epsilon \check{g}_{\epsilon })$ is given by
 \begin{equation}
     (\zeta_\epsilon \check{g}_{\epsilon }) = \begin{Bmatrix}
        \zeta_\epsilon^{R} \hat{g}_{\epsilon }^R & 0 \\
         0 & \zeta_\epsilon^{A} \hat{g}_{\epsilon }^A
     \end{Bmatrix}
 \end{equation}
The matrices $\check{R}_{\Delta }(\epsilon ,\epsilon ^{\prime })
$, $\check{R}_{Q}(\epsilon ,\epsilon ^{\prime })$ are defined as

\begin{eqnarray}
&&\check{R}_{\Delta }(\epsilon ,\epsilon ^{\prime })=\check{g}_{\epsilon
}\delta \check{\Delta}_{\Omega }-\delta \check{\Delta}_{\Omega }\check{g}%
_{\epsilon ^{\prime }}\text{,}  \label{6a} \\
\check{R}_{Q}(\epsilon ,\epsilon ^{\prime }) &=&iD\{\check{g}_{\epsilon }%
\check{Q}_{\nu }\check{g}_{\epsilon +\Omega _{\nu }}\check{Q}_{\mu }-\check{Q}%
_{\nu }\check{g}_{\epsilon +\Omega _{\nu }}\check{Q}_{\mu }\check{g}_{\epsilon
^{\prime }}\}  \label{6b}
\end{eqnarray}
where $\check{Q}_{\nu }=Q_{\nu}\check{\tau}_3$ with 
\begin{equation}
    \check{\tau}_i=\begin{pmatrix}\hat{\tau}_i & 0 \\ 0 & \hat{\tau}_i\end{pmatrix}.
\end{equation}
The order parameter $\check{\Delta}
(t) = \Delta(t) i\check{\tau}_2$ is related to the matrix $\hat{g}^{K}$ via
\begin{equation}
\Delta (t)=\lambda \text{Tr}\int \frac{d\epsilon d\epsilon ^{\prime }}{(2\pi
)^{2}}(-i\hat{\tau}_{2})\hat{g}^{K}(\epsilon ,\epsilon ^{\prime })\exp
(-i\epsilon _{-}t)  \label{Dlt}
\end{equation}
Let us consider first the retarded (advanced) function. The normalisation condition, Eq.\,(\ref{5C}), yields
\begin{equation}
(\hat{g}_{\epsilon }\delta \hat{g}+\delta \hat{g}\hat{g}_{\epsilon ^{\prime
}})^{R(A)}=0  \label{7}
\end{equation}%
Using Eq.\,(\ref{7}),\ we obtain from Eq.\,(\ref{6})
\begin{equation}
\delta \hat{g}^{R(A)}(\epsilon ,\epsilon ^{\prime })=\sum_{\nu ,\mu }\left\{ \frac{
\hat{\varrho}_{\Delta }+\hat{\varrho}_{Q}}{C_{\epsilon ,\epsilon ^{\prime }}}
\right\}^{R(A)}2\pi \delta (\epsilon _{-}+\Omega _{\nu +\mu })\text{,}  \label{8}
\end{equation}%
where $\hat{\varrho}=\hat{g}_{\epsilon }\hat{R}$ and $C_{\epsilon ,\epsilon
^{\prime }}^{R(A)}=(\zeta _{\epsilon }+\zeta _{\epsilon ^{\prime }})^{R(A)}$. The correction $\delta \hat{g}^{K}$ to the Keldysh function can be
represented as a sum of regular and anomalous parts \cite{GorkovEliashberg68}
\begin{equation}
\delta \hat{g}^{K}=\delta \hat{g}^{reg}+\delta\hat{g}^{an}\text{,}  \label{9}
\end{equation}%
The regular part is defined as
\begin{equation}
\delta \hat{g}^{reg}=\delta \hat{g}^{R}t_{\epsilon ^{\prime }}-t_{\epsilon
}\delta \hat{g}^{A}\text{,}  \label{9a}
\end{equation}%
with the matrices $\delta \hat{g}^{R(A)}$ given by Eq.\,(\ref{8}). 
When studying the response of a superconductor to AC radiation, the authors of many works 
used the Matsubara frequency representation. The transition to real energies is carried out with the help of an analytical continuation. 
In our method, we do not use the Matsubara representation and the analytic continuation.

The key point of this approach is to split the Keldysh Green's function $\hat{g}^{K}$ into
a regular $\hat{g}^{reg}$ and anomalous $\hat{g}^{an}$ parts \cite%
{GorkovEliashberg68}. The method of analytical continuation was used, for
example, in a recent paper \cite{SilaevPRB19} (see also Kopnin's book \cite%
{KopninBook01} and references therein), while the former was applied in
Refs. \cite{Artemenko1979,Moor2017}. For the anomalous Green's
function $\delta\hat{g}^{an}$ we find (see Appendix)
\begin{equation}
\delta\hat{g}^{an}=\sum_{\nu ,\mu }\frac{\hat{\varrho}_{\Delta }^{an}+\hat{\varrho}%
_{Q}^{an}}{C_{\epsilon ,\epsilon ^{\prime }}^{an}}2\pi \delta (\epsilon
_{-}+\Omega _{\nu +\mu })  \label{10}
\end{equation}%
with $C_{\epsilon ,\epsilon ^{\prime }}^{an}=\zeta _{\epsilon }^{R}+\zeta
_{\epsilon ^{\prime }}^{A}$. The matrices $\hat{\varrho}_{\Delta }^{an}=%
\hat{g}^{R}\hat{R}_{\Delta }^{an}$ and $\hat{\varrho}_{Q}^{an}=\hat{g}^R\hat{R}^{an}_Q$ are defined as follows
\begin{eqnarray}
\hat{\varrho}_{\Delta }^{an} &=&(\hat{g}_{\epsilon }^{R}\delta \hat{\Delta}%
_{\Omega }\hat{g}_{\epsilon ^{\prime }}^{A}-\delta \hat{\Delta}_{\Omega
})(t_{\epsilon ^{\prime }}-t_{\epsilon })\text{,}  \label{10a} \\
\hat{\varrho}_{Q}^{an} &=&iD[\hat{g}_{\epsilon }^{R}\hat{Q}_{\nu }\hat{g}%
_{\epsilon +\Omega _{\nu }}^{R}\hat{Q}_{\mu }\hat{g}_{\epsilon ^{\prime
}}^{A}-\hat{Q}_{\nu }\hat{g}_{\epsilon +\Omega _{\nu }}^{R}\hat{Q}_{\mu
}](t_{\epsilon'}-t_{\epsilon +\Omega_\nu})  \nonumber \\
&&-iD[\hat{g}_{\epsilon }^{R}\hat{Q}_{\nu }\hat{g}_{\epsilon +\Omega _{\nu
}}^{A}\hat{Q}_{\mu }\hat{g}_{\epsilon ^{\prime }}^{A}-\hat{Q}_{\nu }\hat{g}%
_{\epsilon +\Omega _{\nu }}^{A}\hat{Q}_{\mu }](t_{\epsilon}-t_{\epsilon+\Omega_\nu })\text{,}  \nonumber \\
\end{eqnarray}
where $\epsilon ^{\prime }=\epsilon +\Omega _{\nu +\mu }$. The matrices $%
g_{\epsilon }^{R(A)}=$ $g_{0}^{R(A)}(\epsilon )$ are the Green's functions in the ground state (see Eqs.\,(\ref{4}-\ref{4a})).

Eqns. (\ref{9})-(\ref{10a}) represent one of the main results of our derivation. Using these
equations together with the self-consistency equation (\ref{Dlt}), one can now 
find the function $\delta \hat{\Delta}(t)$ in the second order in $Q_{\nu,\mu}$. The component $\delta \Delta _{\nu \mu }(t)$ at the frequency $%
\Omega _{\nu +\mu }$ is equal to
\begin{widetext}
\begin{equation}
\delta \Delta _{\nu +\mu }(t)=\frac{1}{2}[\delta \Delta (\Omega _{\nu +\mu
})\exp (i\Omega _{\nu +\mu }t)+\delta \Delta (-\Omega _{\nu +\mu })\exp
(-i\Omega _{\nu +\mu }t)]\text{.}  \label{11}
\end{equation}
\end{widetext}
As to the regular part $\delta g^{reg}$, the integral in Eq.\,(\ref{Dlt}) can
be transformed into a sum over Matsubara frequencies ($\epsilon \rightarrow
i\epsilon _{n}$, $\epsilon _{n}=\pi T(2n+1)$) because the retarded
(advanced) Green's functions $\delta g^{R(A)}$ are analytical functions in
the upper (lower) half-plane and the singular points of the regular part $%
\delta g^{reg}$ are determined only by poles of the functions $t_{\epsilon
}=\tanh (\epsilon \beta )$ and $t_{\epsilon ^{\prime }}=\tanh (\epsilon
^{\prime }\beta )$, {\it i.e.} $\epsilon ^{\prime }=i\epsilon _{n}+\Omega _{\nu
+\mu }$ and $\epsilon =i\epsilon _{n}$ with $\epsilon _{n}=\pi T(2n+1)$. The
integral in Eq.\,(\ref{Dlt}) from the anomalous part cannot be reduced to a
sum over Matsubara frequencies and should be calculated explicitly.

\section{Monochromatic Irradiation}

In the case of a monochromatic ac field, it is of interest to calculate the
Fourier components of the $\Delta $ variations: $\delta \Delta _{0}$ and $%
\delta \Delta _{2\Omega }$. This means that the following terms should be
extracted from the sum over frequencies $\Omega _{\nu ,\mu }$:

$\mathcal{A}$) $\Omega _{\nu }\equiv \Omega =-\Omega _{\mu }$,  so that $%
\Omega _{\nu +\mu }\equiv \Omega _{\nu }+\Omega _{\mu }=0$;
and 

$\mathcal{B}$) $\Omega _{\nu }=\Omega =\Omega _{\mu }$, 
so that $\Omega
_{\nu +\mu }\equiv \Omega _{\nu }+\Omega _{\mu }=2\Omega $.

Also the zero Fourier component $\delta \Delta _{0}$ describes the time-averaged
change of $\delta \Delta (t)$ under the influence of an electromagnetic
radiation. The component $\delta \Delta _{2\Omega }$ is the magnitude of the
second harmonic of the amplitude mode excited by the irradiation $Q(t)=Q_{\Omega }\cos (\Omega t)$. It contributes to the third harmonic of the current.

Consider first the case $\mathcal{A}$ ($\Omega _{\nu +\mu }=0$). In this
case, the most interesting quantity is a time-averaged variation of the order parameter, $\langle \delta \Delta (t)\rangle _{t}=\delta \Delta _{0}$, which we discuss in detail below.

\subsection{Eliashberg Effect}

In particular, Eliashberg showed that a microwave irradiation under certain conditions can enhance the order parameter $\Delta $ as well as the
critical temperature $T_{c}$ \cite{Eliashberg70}. This effect as well as the
 enhancement of the critical current was studied in more detail in subsequent works \cite%
{Ivlev71,Ivlev73,Scalapino77,SchmidPRL77,SST-PRB80}. Stability of the nonequilibrium
state with an enhanced $\Delta =\Delta _{eq}+\delta \Delta _{0}$ was
investigated in Ref.\,\cite{ESS80}. Recently, the authors of Ref.\,\cite%
{KlapwijkTikhonov18,SkvortsovRev20} analysed, in particular, the region of
the enhanced $\Delta $ and of the enhanced critical current $j_{c}$ and plot
this region in the plane $(T,\Omega )$. The predicted effect was observed
experimentally, although it was not found to be as strong as expected (see review by
Klapwijk and Visser \cite{KlapwijkRev20} and references therein).

Since in the considered case $\Omega _{\nu +\mu }=0$ we have $\epsilon
=\epsilon ^{\prime }$ and the regular part is given by Eq.\,(\ref{9a}) with $%
t_{\epsilon }\equiv \tanh (\epsilon \beta )=t_{\epsilon ^{\prime }}$
\begin{equation}
\delta \hat{g}^{reg}=t_{\epsilon }(\delta \hat{g}^{R}-\delta \hat{g}^{A})%
\text{,}  \label{E1}
\end{equation}%
where the matrices $\delta \hat{g}^{R(A)}$ are given by Eq.\,(\ref{8}) with $%
\epsilon =\epsilon ^{\prime }$. The matrices $\delta \hat{g}^{R(A)}$ can be
represented in the form
\begin{equation}
\delta \hat{g}^{R(A)}=\frac{\pi }{\zeta _{\epsilon }}[\hat{\tau}%
_{3}(\delta g_{\Delta }+\delta g_{Q})+i\hat{\tau}_{2}(\delta f_{\Delta
}+\delta f_{Q})]^{R(A)} \delta (\epsilon _{-})\text{,}  \label{E2}
\end{equation}%
Here the coefficients, $A$ and $B$, are (for clarity, indices $R(A)$ are
omitted)
\begin{equation}
\delta g_{\Delta }=\delta\Delta B_{\epsilon,\epsilon}\text{, }f_{\Delta }=\delta\Delta A^{(+)}_{\epsilon,\epsilon}\text{,}  \label{E2a}
\end{equation}
\begin{eqnarray}
\delta g_{Q} &=&iDQ^2(A^{(-)}_{\epsilon,\epsilon}g_{\epsilon +\Omega }-B_{\epsilon,\epsilon}f_{\epsilon +\Omega })\text{, }
\label{E2b} \\
\delta f_{Q} &=&-iDQ^2(A^{(+)}_{\epsilon,\epsilon}f_{\epsilon +\Omega }+B_{\epsilon,\epsilon}g_{\epsilon +\Omega })\text{.%
}  \label{E2B}
\end{eqnarray}
The coefficients $A^{(\pm )}$ and $B$ are defined as
\begin{eqnarray}
A^{(\pm )}_{\epsilon ,\epsilon ^{\prime }} &=&1\pm (g_{\epsilon }g_{\epsilon
^{\prime }}+f_{\epsilon }f_{\epsilon ^{\prime }})  \label{E3} \\
\text{ }B_{\epsilon ,\epsilon ^{\prime }} &=&g_{\epsilon }f_{\epsilon
^{\prime }}+f_{\epsilon }g_{\epsilon ^{\prime }}\text{, }  \label{E3a} \\
C_{\epsilon ,\epsilon ^{\prime }} &=&\zeta _{\epsilon }+\zeta _{\epsilon
^{\prime }}\text{,}  \label{E3b}
\end{eqnarray}%
with $\epsilon =\epsilon ^{\prime }$ in this particular case. The anomalous
function $\delta\hat{g}^{an}$ is given by Eq.\,(\ref{10}) with $\hat{\varrho}_{\Delta }^{an}=0$
and $\delta f_{Q}^{an}(\epsilon ,\epsilon ^{\prime })$ equal to
\begin{widetext}    
\begin{equation}
\delta f_{Q}^{an}(\epsilon ,\epsilon ^{\prime })=-iDQ^2(t_{\epsilon +\Omega }-t_{\epsilon
})\frac{(f_{\epsilon +\Omega }^{R}-f_{\epsilon +\Omega
}^{A})A^{an(+)}_{\epsilon, \epsilon}+(g_{\epsilon +\Omega }^{R}-g_{\epsilon +\Omega }^{A})B^{an}_{\epsilon, \epsilon}}%
{C_{\epsilon, \epsilon}^{an}}\text{,}  \label{E4}
\end{equation}%
\end{widetext}
with $A^{an(\pm)}_{\epsilon, \epsilon'}=1\pm(g_{\epsilon }^{R}g_{\epsilon' }^{A}+f_{\epsilon
}^{R}f_{\epsilon'}^{A})$, $B^{an}_{\epsilon,\epsilon'}=g_{\epsilon }^{R}f_{\epsilon
'}^{A}+f_{\epsilon }^{R}g_{\epsilon' }^{A}$, $C_{\epsilon, \epsilon' }^{an}=\zeta
_{\epsilon }^{R}+\zeta _{\epsilon'}^{A}$ and $\epsilon'=\epsilon$.

Let us now find the variation of $\Delta $. In the considered case, the
self-consistency equation given by Eq.\,(\ref{Dlt}) is
\begin{equation}
\delta \Delta (t)=-i\lambda \int \frac{d\bar{\epsilon}}{2\pi }\int \frac{%
d\epsilon _{-}}{2\pi }\text{Tr}\hat{\tau}_{2}\{\delta \hat{g}^{reg}+\delta
\hat{g}^{an}\}\exp (i\epsilon _{-}t)  \label{E5}
\end{equation}%
For the $\delta \Delta (t)$ averaged over time we obtain
\begin{equation}
\delta \Delta =-i\lambda \int \frac{d\epsilon }{2\pi }\text{Tr}\hat{\tau}%
_{2}\{\delta \hat{g}^{reg}+\delta \hat{g}^{an}\}|_{\epsilon _{-}=0}\text{.}
\label{E5a}
\end{equation}%
This equation can be written in the form (see Appendix)
\begin{eqnarray}\label{Eq:PhaseGapEE}
\lefteqn{4T\delta \Delta _{0}\Delta ^{2}\sum_{n}\frac{1}{\zeta _{n}^{3}}  = } && \nonumber \\
&&-DQ^2\left[ T\Delta 
\text{Re}\sum_{n}\frac{\omega (2\omega +i\Omega )}{\zeta _{\omega }^{3}\zeta
    _{\omega +i\Omega }} 
    +\int \frac{d\epsilon }{2\pi }\delta f_{Q}^{an}\right]
\end{eqnarray}
In Fig.2 we plot the dependence of $\delta \Delta _{0}$ (in the normalized
form $\delta \Delta _{0}/DQ^{2}$) as a function of $\Omega $ and $T$ for a given $\gamma$. We also
show the region in the ($T,\Omega $) plane, where $\delta \Delta _{0}$ is
positive, i. e. one finds a stimulation of superconductivity especially in a vicinity of the superconducting transition temperature. Furthermore, smaller $\gamma$ enlarges the region of stimulated superconductivity. The results are consistent with those obtained previously in Refs. (\cite{KlapwijkTikhonov18,SkvortsovRev20}).
\begin{figure} [htbp]
    \centering
    \includegraphics[width=0.43 \textwidth]{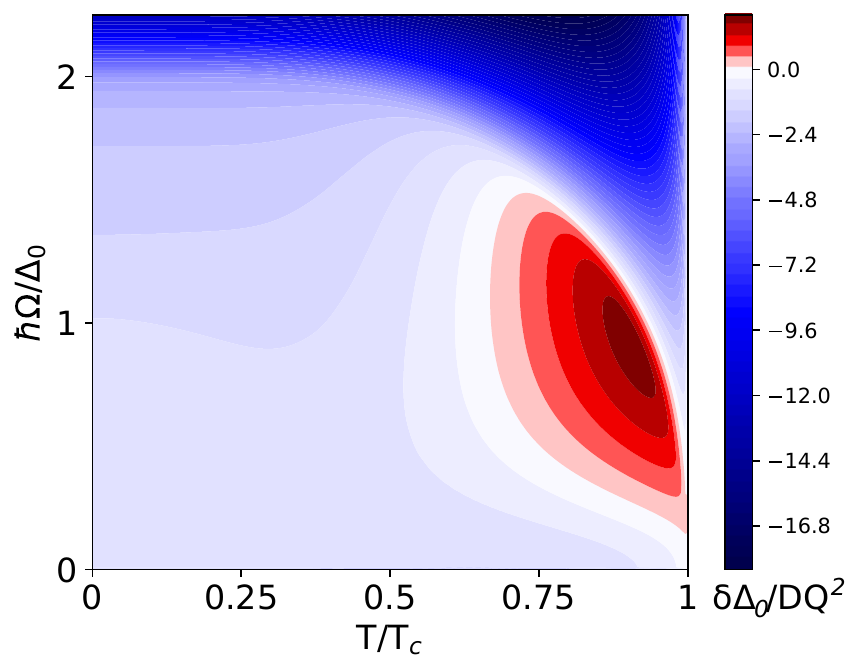}
    \caption{Calculated time-averaged
    gap correction $\delta\Delta_0$ using Eq.\,(\ref{Eq:PhaseGapEE}) for $\gamma=0.02\Delta_0$. Note, we set the polarization operator given by the factor of the left-hand side to $1$, as it's solely positive and real. The area in red represents the area of stimulated (enhanced) superconductivity.}
    \label{fig:phase_Eliashberg}
\end{figure}

\subsection{Second Harmonic of $\protect\delta \Delta $.}

\begin{figure} [htbp]
    \centering
    \includegraphics[width=0.49 \textwidth]{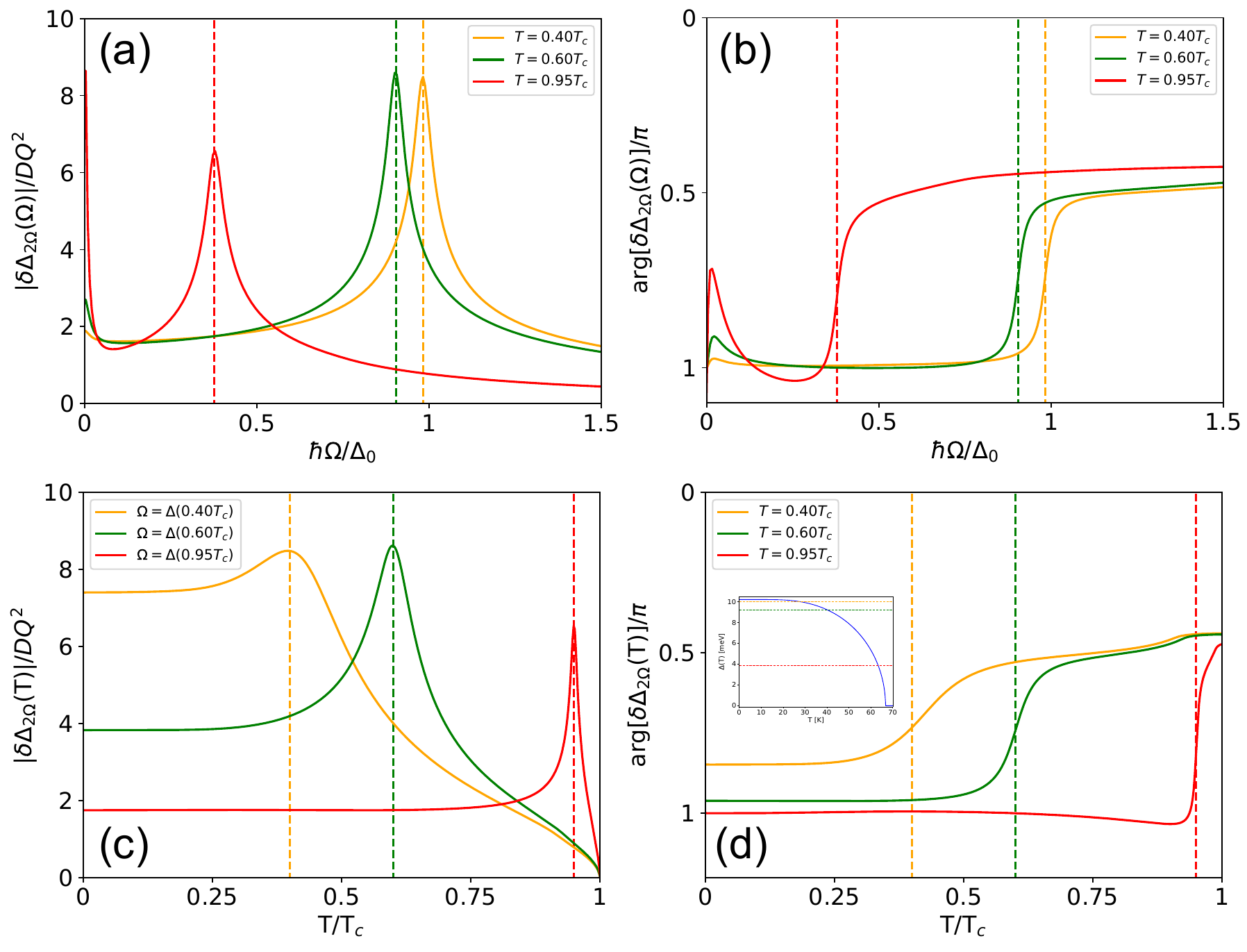}
    \caption{Calculated frequency dependence of the amplitude (a) and the phase (b) of the gap oscillation $\delta\Delta_{2\Omega}$. The panels (c) and (d) show a corresponding temperature dependence of the gap oscillation $\delta\Delta_{2\Omega}$. We observe a peak accompanied by a phase shift at the resonance condition $\Omega=\Delta(T)$, which is highlighted by the dashed lines. The inset in (d) shows the equilibrium gap $\Delta (T)$ as a function of the temperature.}
    \label{fig:GapOsc_Graph}
\end{figure}

Let us now consider the case $\mathcal{B}$ and set $\epsilon ^{\prime }=\epsilon +2\Omega $. The variation of $\Delta $ can be represented as follows
\begin{equation}
\delta \Delta _{2\Omega }=\frac{\delta \Delta _{Q}^{reg}(2\Omega )+\delta
\Delta _{Q}^{an}(2\Omega )}{P_{\Delta }^{reg}(2\Omega )+P_{\Delta
}^{an}(2\Omega )}\text{,}  \label{S1}
\end{equation}%
Here,
\begin{eqnarray}
P_{\Delta }^{an}(2\Omega ) &=&-2\sinh (2\Omega \beta )\int \frac{d\epsilon }{%
2\pi }r_{\epsilon }\frac{A^{an(+)}_{\epsilon, \epsilon+2\Omega}}{C_{\epsilon ,\epsilon +2\Omega
}^{an}}\text{,}  \label{S2} \\
P_{\Delta }^{reg}(2\Omega ) &=&4T\sum_{n}\left(\frac{{A}_{n, n+i2\Omega}^{(+)}}{{C%
}_{n, n+i2\Omega}}-\frac{1}{\zeta _{n}}\right)  \nonumber \\
&=&-4Ti\sum_{n}\frac{\Delta ^{2} }{\zeta _{n}^{2}\left(\zeta
_{n}+\zeta _{n+i2\Omega }\right)}
\end{eqnarray}%
where $\zeta _{n}=\sqrt{\epsilon _{n}^{2}+\Delta ^{2}}$ and the subindex $n$ denotes in $A^{(\pm)}_n$, $B_n$ and $C_n$ that the functions are expressed in terms of the Matsubara frequencies $\epsilon=i\epsilon_n=i\pi T(2n+1)$. The terms in the numerator are
\begin{equation}
\delta \Delta _{Q}^{reg}=2TDQ^{2}\text{Re}\sum_{n\geqslant 0}\frac{%
f_{n+i\Omega}A_{n, n+i2\Omega}^{(+)}+g_{n+i\Omega}B^{}_{n, n+i2\Omega}}{\zeta
_{n}+\zeta _{n+i2\Omega }}  \label{S3}
\end{equation}%
and
\begin{eqnarray}
&&\delta\Delta_{Q}^{an}=-iDQ^{2}\sinh (2\Omega \beta )\int \frac{%
d\epsilon }{2\pi }\frac{r_{\epsilon }}{C^{an}_{\epsilon _{-},\epsilon _{+}}} \times \nonumber \\ && \left(A^{an(+)}_{\epsilon _{-},\epsilon _{+}}\left[ F^{(+)}-F^{(-)}\tanh (\epsilon \beta
)\tanh (\Omega \beta ) \right] \right. \\
&& \left. +B^{an}_{\epsilon _{-},\epsilon _{+}}\left[G^{(+)}-G^{(-)}\tanh (\epsilon \beta
)\tanh (\Omega \beta )\right]\right)\text{.}\nonumber
\end{eqnarray}
The temperature factor $r(\epsilon ,\Omega )$ is defined as
\begin{equation}
r_{\epsilon } =\left[\cosh (\epsilon _{+}\beta )\cosh (\epsilon _{-}\beta
)\right]^{-1}\text{,}  \label{S5} \\
\end{equation}%
%
where $\epsilon _{\pm }=\epsilon \pm \Omega $. The functions $F^{(\pm )}$, $%
G^{(\pm )}$ are given by
\begin{equation}
F^{(\pm )}=f_{\epsilon }^{R}\pm f_{\epsilon }^{A}\text{, }G^{(\pm
)}=g_{\epsilon }^{R}\pm g_{\epsilon }^{A}\text{,}  \label{S6}
\end{equation}
At low frequencies, $\Omega \Rightarrow 0$, we obtain $\delta \Delta
_{Q}^{an}\Rightarrow 0$ and $\delta \Delta _{Q}^{reg}$ has the same form as in a static case
\begin{equation}
\frac{\delta \Delta _{Q}^{reg}}{P_\Delta^{reg}}=-DQ^{2}\frac{\sum_{n\geqslant 0}\frac{\epsilon _{n}^{2}}{%
\zeta _{n}^{4}}}{\sum_{n\geqslant 0}\frac{\Delta }{\zeta _{n}^{3}}}  \label{S7}
\end{equation}
In Fig.\,\ref{fig:GapOsc_Graph} we plot the frequency dependence of the normalized variation of the
second harmonic of $\Delta$,  $\delta \tilde{\Delta}_{2\Omega }\equiv $ $\delta
\Delta _{2\Omega }/DQ^2$ and $\Delta_0$ refers to the equilibrium $T=0$ value of the superconducting gap. As expected the amplitude mode gives a resonant contribution around the corresponding value of $\Delta(T)$ with a characteristic phase shift of $\pi/2$.

\section{The Current induced by ac Field}

To investigate the induced current, we consider an ac electric field incident on a superconducting film of the
thickness $2d$. The field consists of a probe $E_{\omega }$ and pump $E_{\Omega }$ field components: $E(t)=E_{\omega }\cos (\omega t)+E_{\Omega }\cos
(\Omega t)$. This field excites alternating currents $j(t)$ of various
harmonics. In the third order in the amplitude $E$, the induced current consists of harmonics with frequencies $\omega $, and $2\Omega \pm \omega $.

The current $j(t)$ is determined by 
\begin{equation}
\pmb{j}(t)=i\frac{\sigma_0 }{4e}\text{Tr} \left[ \hat{\tau}_{3}\int d\tau \{\hat{g}[%
\mathbf{\hat{Q},}\hat{g}]\exp (i\Omega \tau )+[\mathbf{\hat{q},}\hat{g}%
\mathbf{]}\exp (i\omega \tau )\}_{3}^{K}  \right]
\label{Cr1}
\end{equation}%
where $\mathbf{\hat{Q}=Q}\hat{\tau}_{3}$, the momentum $\mathbf{Q}$ is the
gauge-invariant quantity related to the phase of the superconductor $\chi $
and the vector potential $\mathbf{A}$: $\mathbf{Q=\nabla }\chi -(e/c)\mathbf{%
A}$. In the case of $\chi =0$, the vector potential $\mathbf{A}$ relates to
the electric field as
\begin{equation}
\mathbf{E}_{Q}(t)=\frac{\hbar i\Omega }{e}\mathbf{Q}\text{, }\mathbf{E}%
_{q}(t)=\frac{\hbar i\omega }{e}\mathbf{q}\text{. }  \label{Cr2}
\end{equation}%
where $\omega $ and $\Omega $ is a probe (signal) and a pump frequency. These
frequencies may in general coincide: $\omega =\Omega $.

We decompose all functions to the power of $\mathbf{Q}$ considering the
quantity $DQ^{2}/\Delta(T)\,$\ as a small parameter. In the lowest order we
obtain a linear response to the action of two signals with $\Omega $ and $%
\omega $. It equals
\begin{eqnarray}
\pmb{j}_{1}(t)&=&i\frac{\sigma_0 }{4e}\text{ }\int \frac{d\epsilon }{2\pi }\left[%
\mathbf{Q}\text{ }\exp (i\Omega t\mathbf{)}\{\hat{g}_{\epsilon -\Omega }%
\mathbf{\langle }\hat{g}_{\epsilon })\rangle \}_{0}^{K}+ \right.
\nonumber \\
&&\left.+\text{ }\mathbf{q}%
\text{ }\exp (i\omega t\mathbf{)}\{\hat{g}_{\epsilon -\omega }\mathbf{%
\langle }\hat{g}_{\epsilon }\rangle \}_{0}^{K}\right]\text{.} \label{Cr3aa} 
\end{eqnarray}%
Here $\mathbf{\langle }\hat{g}\rangle =\hat{\tau}_{3}\hat{g}\hat{\tau}_{3}$,
$\{...\}_{0}=(1/2)$Tr$\{...\}$ and $\hat{g}_{\epsilon -\omega }^{R}\equiv
\hat{g}_{0}^{R}(\epsilon -\omega )$ etc. The current $\pmb{j}_{1q}(t)\sim
\mathbf{q}$ $\exp (i\omega t\mathbf{)}$ can be represented as
\begin{equation}
\pmb{j}_{1q}(t)=i\frac{\sigma_0 }{4e}\text{ }\mathbf{q}\text{ }\exp
(i\omega t\mathbf{)[}J_{1}^{reg}+J_{1}^{an}\mathbf{]}\text{,}  \label{Cr3a}
\end{equation}%
with
\begin{eqnarray}
J_{1}^{reg} &=&\int \frac{d\epsilon }{2\pi }\{\hat{g}_{\epsilon }^{R}\mathbf{\langle }\hat{g}_{\epsilon
+\omega }^{R}\rangle t_{\epsilon +\omega }-\hat{g}_{\epsilon }^{A}\mathbf{%
\langle }\hat{g}_{\epsilon +\omega }^{A}\rangle t_{\epsilon }\}_{0}
\label{Cr3b} \\
J_{1}^{an} &=&-\int \frac{d\epsilon }{2\pi }\{\hat{g}_{\epsilon }^{R}\mathbf{\langle }\hat{g}_{\epsilon
+\omega }^{A}\rangle \}_{0}(t_{\epsilon +\omega }-t_{\epsilon })\text{,}
\label{Cr3c}
\end{eqnarray}
The next correction\ to the first $\omega $ harmonic in the presence of a
pump is
\begin{eqnarray}
\delta\pmb{j}(t) &=&i\frac{\sigma_0 }{4e}\mathbf{q}\int d\tau \exp
(i\omega \tau )\{\delta \hat{g}(t,\tau )\langle \hat{g}_{0}(\tau -t)\rangle +
\nonumber \\
&&+\langle \hat{g}_{0}(t-\tau )\rangle \delta \hat{g}(\tau ,t)\}_{0}^{K}%
\text{.}  \label{Cr4}
\end{eqnarray}%
In the Fourier representation, the current acquires the form
\begin{eqnarray}
\lefteqn{\delta \pmb{j}_{\omega }(t) =i\frac{\sigma_0 }{4e}\mathbf{q}_{\omega
}\sum_{\omega,\nu ,\mu }\exp ((i\omega +\Omega _{\nu +\mu })t)\times}
 \\
&&\int \frac{d\epsilon }{2\pi }\{\delta \hat{g}_{\epsilon, \epsilon+\Omega_{\nu+\mu} }\langle
\hat{g}_{\epsilon +\omega +\Omega _{\nu +\mu }}\rangle +\langle \hat{g}%
_{\epsilon -\omega }\rangle \delta \hat{g}_{\epsilon+\Omega_{\nu+\mu}}\}_{0}^{K}\nonumber\text{, }
\end{eqnarray}%
Here $\delta \hat{g}_{\epsilon }$ contains the terms of the order of $Q^{2}$%
, and the summation is performed for $\pm \omega $, $\Omega _{\nu }=\pm
\Omega _{\mu }=\pm \Omega $. Therefore the current $\delta \pmb{j}%
_{\omega }(t)$ contains harmonics $\delta \pmb{j}_{\omega ,\Omega }\infty
\delta \pmb{j}_{\omega ,0}(Q^{2})\exp (i\omega t)$, $\delta \pmb{j}%
_{\omega ,3}(Q^{2})\exp (i(2\Omega \pm \omega )t)$. The first harmonic
corresponds to $\Omega _{\nu }=-\Omega _{\mu }=\Omega $, whereas the third
harmonic corresponds to $\Omega _{\nu }=\Omega _{\mu }=\Omega $. In
obtaining this equation, we used the relation $\delta \hat{g}_{\epsilon
,\epsilon ^{\prime }}=\delta \hat{g}_{\epsilon }2\pi \delta (\epsilon
^{\prime }-\epsilon \pm \Omega _{\nu +\mu })$.

The correction $\delta \hat{g}_{\epsilon ,\epsilon
^{\prime }}$ consists of all second order combinations of $\mathbf{Q}$ and $\mathbf{q}$ and we took into account that $\mathbf{Q}\perp\mathbf{q}$ and $|\mathbf{q}|\ll|\mathbf{Q}|$ which leaves only the term  $\propto Q^2$.
%
%
First, we consider the correction to the conductance caused by the Eliashberg effect (EE).

\subsection{Correction to the Current due to EE.}

In this case, the frequency $\Omega _{\nu +\mu }=\Omega _{\nu }+\Omega _{\mu
}=0$ and the correction $\delta \hat{g}_{\epsilon ,\epsilon
^{\prime }}$ can be written as follows
\begin{equation}
\delta \hat{g}_{\epsilon ,\epsilon ^{\prime }}=(DQ^{2})\delta \hat{g}%
_{\epsilon}2\pi \delta (\epsilon -\epsilon ^{\prime })+(-\Omega ).
\label{CrE1}
\end{equation}%
Then, the integral Eq.\,(54) acquires the form
\begin{equation}
 \pmb{j}^{EE}(t)=i\frac{\sigma_0 }{4e}\mathbf{q}_{\omega }\exp
(i\omega t)J^{EE}\text{,}  \label{CrE2}
\end{equation}%
with
\begin{equation}
J^{EE}=\int \frac{d\epsilon }{2\pi }\{\delta \hat{g}_{\epsilon
}^{R}\langle \hat{g}_{\epsilon +\omega }^{K}\rangle +\delta \hat{g}%
_{\epsilon }^{K}\langle \hat{g}_{\epsilon +\omega }^{A}\rangle +\langle
\hat{g}_{\epsilon -\omega}^{R}\rangle \delta \hat{g}_{\epsilon }^{K}+\langle \hat{g}%
_{\epsilon -\omega }^{K}\rangle \delta \hat{g}_{\epsilon }^{A}\}_{0}\text{,}
\label{CrE3}
\end{equation}%
The Keldysh component $\delta \hat{g}_{\epsilon }^{K}$ is defined as
\begin{eqnarray}
\delta \hat{g}_{\epsilon }^{K} &=&\delta \hat{g}_{\epsilon }^{reg}+\delta
\hat{g}_{\epsilon }^{an}\text{,}  \label{CrE4} \\
\delta \hat{g}_{\epsilon }^{reg} &=&t_{\epsilon }(\delta \hat{g}_{\epsilon
}^{R}-\delta \hat{g}_{\epsilon }^{A})  \label{CrE4a}
\end{eqnarray}%
The integral $J^{EE}$ consists now of three terms
\begin{equation}
J^{EE}=J^{reg}+J^{AN}+J^{an}\text{.}  \label{CrE5}
\end{equation}%
The regular part is
\begin{widetext}
    
\begin{eqnarray}
J^{reg} &=&\int \frac{d\epsilon }{2\pi }\{\delta \hat{g}_{\epsilon
}^{R}[\langle \hat{g}_{\epsilon +\omega }\rangle t_{\epsilon +\omega
}+\langle \hat{g}_{\epsilon -\omega }\rangle t_{\epsilon }]^{R}-\delta \hat{g}_{\epsilon }^{A}[\langle \hat{g}_{\epsilon +\omega }\rangle
t_{\epsilon }+\langle \hat{g}_{\epsilon -\omega }\rangle t_{\epsilon -\omega
}]^{A}\}_{0}\text{, }  \label{CrE6}
\end{eqnarray}%
The anomalous terms $J^{AN}$ and $J^{an}$ are equal to
\begin{eqnarray}
J^{AN} &=&-\int \frac{d\epsilon }{2\pi }\{\delta \hat{g}_{\epsilon
}^{R}\langle \hat{g}_{\epsilon +\omega }^{A}\rangle (t_{\epsilon +
\omega}-t_{\epsilon })+\delta \hat{g}_{\epsilon }^{A}\langle \hat{g}_{\epsilon
-\omega }^{R}\rangle (t_{\epsilon }-t_{\epsilon -\omega })\}_{0}\text{,}
\label{CrE6A} \\
\text{ }J^{an} &=&\int \frac{d\epsilon }{2\pi }\{\delta \hat{g}_{\epsilon
}^{an}\langle \hat{g}_{\epsilon -\omega }^{R}+\hat{g}_{\epsilon +\omega
}^{A}\rangle \}_{0}\text{.}  \label{CrE6B}
\end{eqnarray}
Here, $\delta \hat{g}_{\epsilon }^{R(A)}$ matrix is 
\begin{eqnarray}
\delta \hat{g}_{\epsilon }^{R(A)}
&=& \frac{1}{2\zeta _{\epsilon }^{R(A)}}\{\hat{\tau}_3(\delta g_\Delta + \delta g_Q)
+i\hat{\tau}_2(\delta f_\Delta + \delta f_Q)\}^{R(A)}
\end{eqnarray}

\end{widetext}
The anomalous function $\delta \hat{g}_{\epsilon }^{an}$ is
\begin{eqnarray}
\delta \hat{g}_{\epsilon }^{an} 
&=&iDQ^2\frac{t_{\epsilon +\Omega }-t_{\epsilon }}{(\zeta _{\epsilon
}^{R}+\zeta _{\epsilon }^{A})}\{N_{3}^{an}\hat{\tau}_{3}-i\hat{\tau}%
_{2}N_{2}^{an}\},
\end{eqnarray}%
where $N^{an}_{2, 3}$ are
\begin{eqnarray}
N_{3}^{an} &=&G_{\epsilon+\Omega}^{(-)}A_{\epsilon,\epsilon }^{an(-)}-F_{
\epsilon+\Omega}^{(-)}B^{an}_{\epsilon,\epsilon} \label{CrE8a} \\
N_{2}^{an} &=&F_{\epsilon+\Omega}^{(-)}A_{\epsilon, \epsilon}^{an(+)}+G_{{%
\epsilon+\Omega}}^{(-)}B^{an}_{\epsilon, \epsilon}\text{,}  \label{CrE8B} 
\end{eqnarray}
From the current, we can derive the correction to the complex conductivity 
\begin{equation}
    \delta\sigma(\omega, \Omega)=\frac{\sigma_0}{4\hbar\omega}J^{EE}(\omega, \Omega),
\end{equation}
while the equilibrium conductivity is given by:
\begin{equation}
    \sigma^{(1)}(\omega)=\frac{\sigma_0}{4\hbar\omega}[J_1^{reg}+J_1^{an}]
\end{equation}

\begin{figure} [htbp]
    \centering
    \includegraphics[width=0.49 \textwidth]{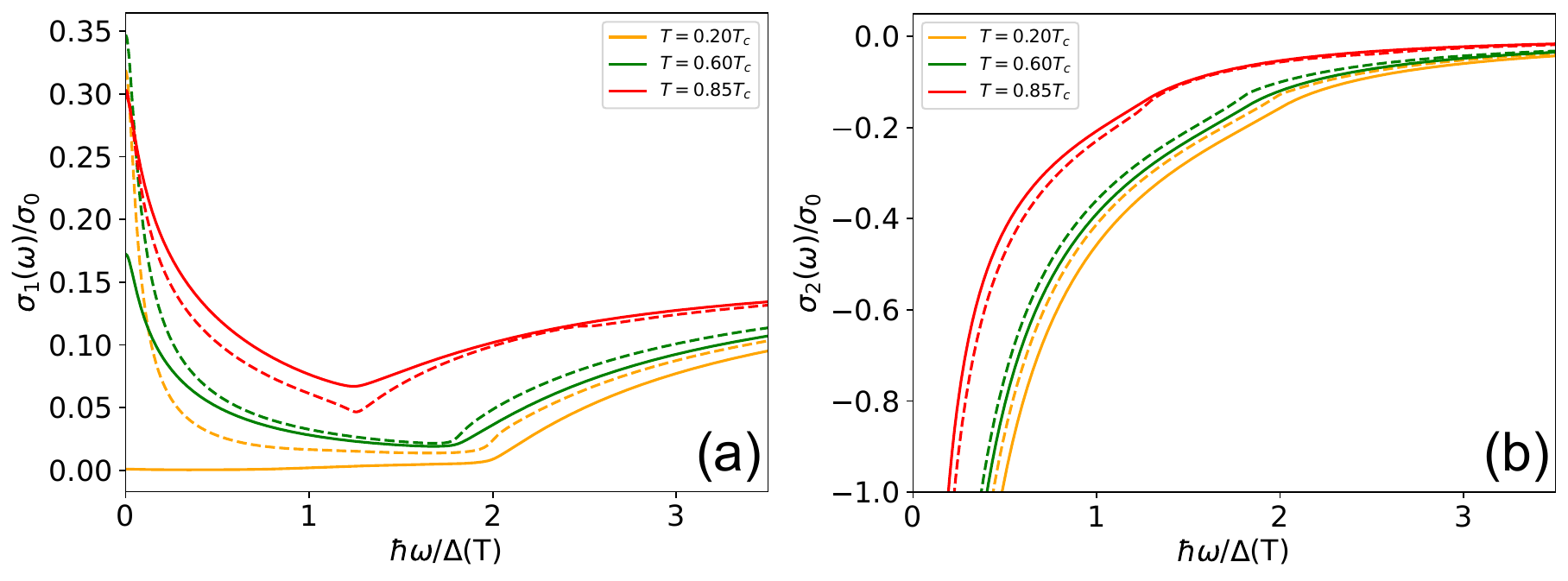}
    \caption{Calculated photo-excited conductivity  $\sigma(\omega,\Omega)$. The real part (a) of the conductivity $\sigma_1(\omega)$ and the imaginary part (b) $\sigma_2(\omega)$ are shown as function of the probe frequency $\omega$. The dashed (solid) curves represent the photo-excited (equilibrium) conductivity. The calculations were done for $DQ^2=0.005\Delta(T)$. The pump-frequency is fixed at  $\Omega=2\Delta(T)$.}
    \label{fig:Graph_sigma_frequency}
\end{figure}

In Fig.\,\ref{fig:Graph_sigma_frequency} we show the total conductivity $\sigma (\omega,\Omega)=\sigma^{(1)}(\omega)+\delta\sigma(\omega, \Omega)$ (real and imaginary part) as a function of the probe frequency, $\omega$, in comparison to its equilibrium behavior, $\sigma^{(1)}(\omega)$, for several temperatures and pump frequency of $\Omega=2\Delta(T)$. Observe that within the linear response and low temperatures $T=0.2T_c$ the optical conductivity is gapped approximately to 2$\Delta(T)$ and its evolution with temperature follow the standard behavior, dictated by the thermal excitation of quasiparticles in equilibrium superconductor, see for example Ref.\,\cite{Boyack2023}. The effect of the pump at low temperature ($T=0.2T_c$) an be viewed as the effect of the "effective temperature" as $\sigma_1(\omega)$ become gapless and increases at lower $\omega$ due to pair-breaking effect of the pump pulse and resulting formation of the quasiparticles in a similar fashion as the increasing temperature would do. The situation, however, changes for higher temperatures and especially for $T=0.85T_c$ where one observes clearly the Eliashberg effect in which $\sigma_1(\omega)$ shows much stronger gap features than it is in the equilibrium.  

\begin{figure} [htbp]
    \centering
    \vspace{3mm}
    \includegraphics[width=0.49 \textwidth]{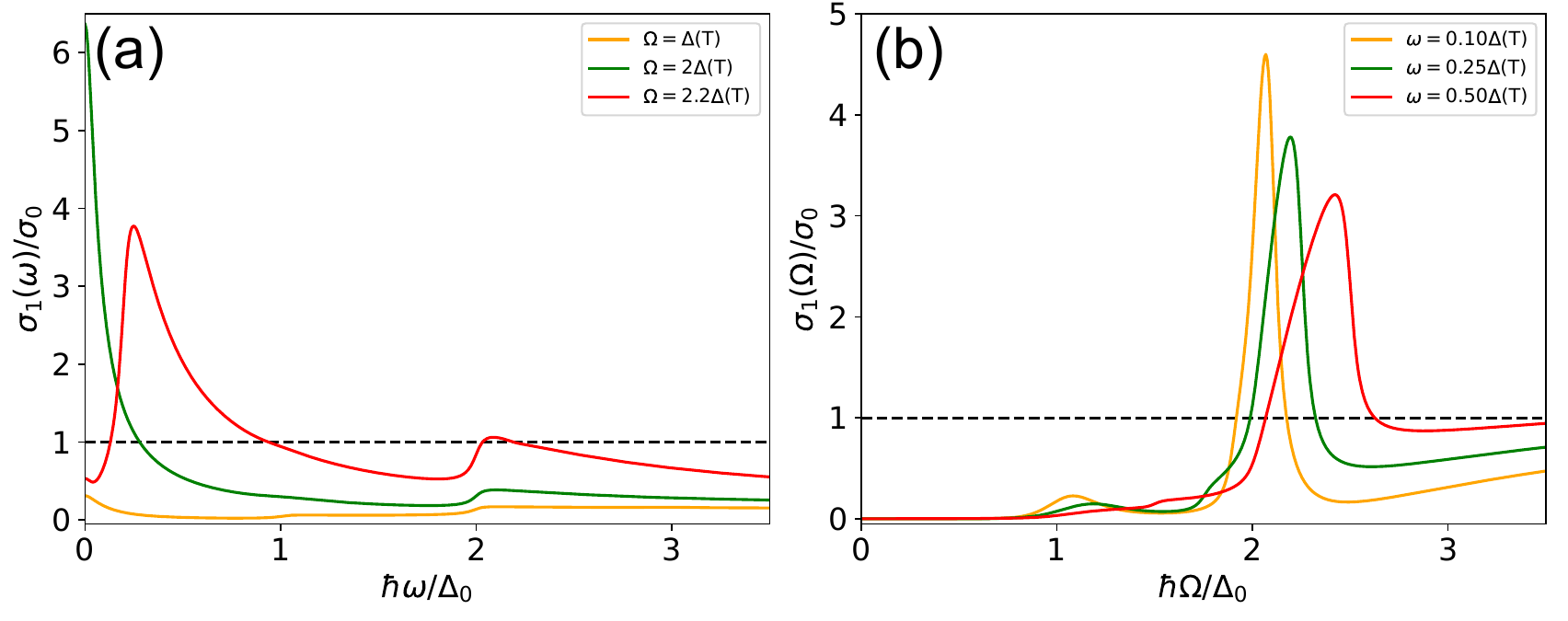}
       \caption{Calculated real part of the photo-excited complex conductivity, $\sigma (\omega, \Omega)$. In (a) $\sigma_1(\omega,\Omega)$ is shown as function of the probe-frequency $\omega$ for various pump-frequencies $\Omega$. In (b) the real part of the conductivity $\sigma_1(\omega,\Omega)$ is shown as a function of the pump-frequency $\Omega$ for various fixed probe-frequencies $\omega$. The calculation were done for $DQ^2=0.1\Delta(T)$ at $T=0.1T_c$.}
    \label{fig:Graph_sigma1_frequency}
\end{figure}

In Fig.\,\ref{fig:Graph_sigma1_frequency} we show the behavior of $\sigma(\omega,\Omega)$ at a low temperature of $T=0.1T_c$ as a function of the probe (a) and the pump frequency (b). One could clearly see that effect of the pump arises for $\Omega \geq \Delta(T)$ and is the most prominent one for $\Omega \sim 2\Delta(T)$. Note that we plot the curves in Figs. (4-5) for particular magnitudes of ratio $p=DQ^{2}/\Delta$ which is assumed to be small. But all the functions (Corrections of the Green’s functions and variation of $\Delta$) are proportional to this parameter. Therefore, an increase of $p$ means stretching the graphs along the y-axis.  

\subsection{Amplitude of the third Harmonic.}

Here, we consider an amplitude of the third harmonic of the form $
j^{TH}(t)\infty j^{TH}\exp (i(2\Omega \pm \omega )$. In this case, the frequency $\Omega _{\nu +\mu }=\Omega _{\nu }+\Omega _{\mu
}=2\Omega $ and  the correction $\delta \hat{g}_{\epsilon ,\epsilon
^{\prime }}$ can be written as follows
\begin{equation}
\delta \hat{g}_{\epsilon ,\epsilon ^{\prime }}=(DQ^{2})\delta \hat{g}%
_{\epsilon}2\pi \delta (\epsilon _{-}+2\Omega)%
\text{.}  \label{CrT1}
\end{equation}
The current in Eq.\,(54) is given by

\begin{equation}
 \pmb{j}^{TH}(t)=i\frac{\sigma }{4e}\mathbf{q}_{\omega }\exp
(i(2\Omega \pm \omega )t)J^{TH}\text{,}  \label{CrT2}
\end{equation}%
with $J^{TH}$ equal to

\begin{equation}
J^{TH}=\tilde{J}^{reg}+\tilde{J}^{AN}+\tilde{J}^{an}\text{.}  \label{CrT3}
\end{equation}

The currents $\tilde{J}^{reg}$, $\tilde{J}^{AN}$, and $\tilde{J}^{an}$ are
defined as follows

\begin{widetext}

\begin{eqnarray}
\tilde{J}^{reg} &=&\int \frac{d\epsilon }{2\pi }\{\delta \hat{g}_{\epsilon
}^{R}[\langle \hat{g}_{\tilde{\epsilon}}\rangle t_{\tilde{\epsilon}}+\langle
\hat{g}_{\epsilon -\omega }\rangle t_{\epsilon +2\Omega }]^{R} -\delta \hat{g}_{\epsilon }^{A}[\langle \hat{g}_{\tilde{\epsilon}}\rangle
t_{\epsilon }+\langle \hat{g}_{\epsilon -\omega }\rangle t_{\epsilon -\omega
}]^{A}\}_{0}\text{, }  \label{CrT4} \\
\tilde{J}^{AN} &=&-\int \frac{d\epsilon }{2\pi }\{\delta \hat{g}_{\epsilon
}^{R}\langle \hat{g}_{\tilde{\epsilon}}^{A}\rangle (t_{\tilde{\epsilon}%
}-t_{\epsilon +2\Omega })+\delta \hat{g}_{\epsilon }^{A}\langle \hat{g}%
_{\epsilon -\omega }^{R}\rangle (t_{\epsilon }-t_{\epsilon -\omega })\}_{0}%
\text{,}   \label{CrT4a}\\
\tilde{J}^{an} &=&\int \frac{d\epsilon }{2\pi }\{\delta \hat{g}_{\epsilon
}^{an}\langle \hat{g}_{\epsilon -\omega }^{R}+\hat{g}_{\tilde{\epsilon}%
}^{A}\rangle \}_{0}  \label{CrT4b}
\end{eqnarray}
where $\tilde{\epsilon}=2\Omega \pm \omega $ and the matrix $\delta\hat{g}^{R(A)}_\epsilon$ is given by:
\begin{equation}
    \delta \hat{g}_{\epsilon }^{R(A)} =\frac{1}{\zeta _{\epsilon }^{R(A)}+\zeta _{\epsilon+2\Omega}^{R(A)}}\{\delta
    \hat{\Delta}_{2\Omega}-\hat{g}_{\epsilon }^{R}\delta \hat{\Delta}_{2\Omega}\hat{g}_{\epsilon+2\Omega
    }^{R}+iDQ^2[\langle \hat{g}_{\epsilon +\Omega }\rangle -\hat{g}_{\epsilon
    }\langle \hat{g}_{\epsilon +\Omega }\rangle \hat{g}_{\epsilon +2\Omega}]\}^{R(A)} 
\end{equation}
The anomalous function $\delta\hat{g}^{an}_\epsilon$ is
\begin{equation}
\begin{split}
    \delta \hat{g}_{\epsilon }^{an} =\frac{1}{\zeta _{\epsilon }^{R}+\zeta _{\epsilon+2\Omega }^{A}}\{[\delta
    \hat{\Delta}_{2\Omega}-\hat{g}_{\epsilon }^{R}\delta \hat{\Delta}_{2\Omega}\hat{g}_{\epsilon+2\Omega
    }^{A}]&(t_\epsilon-t_{\epsilon+2\Omega})+iDQ^2[\langle \hat{g}^R_{\epsilon +\Omega }\rangle -\hat{g}_{\epsilon
    }^R\langle \hat{g}_{\epsilon +\Omega }^R\rangle \hat{g}^A_{\epsilon +2\Omega}](t_{\epsilon +\Omega}-t_{\epsilon+2\Omega})\\
    &+iDQ^2[\langle \hat{g}^A_{\epsilon +\Omega }\rangle -\hat{g}_{\epsilon
    }^R\langle \hat{g}_{\epsilon +\Omega }^A\rangle \hat{g}^A_{\epsilon +2\Omega}](t_{\epsilon}-t_{\epsilon+\Omega})\}
\end{split}
\end{equation}
\end{widetext}

\begin{figure} [htbp]
    \centering
    \includegraphics[width=0.49 \textwidth]{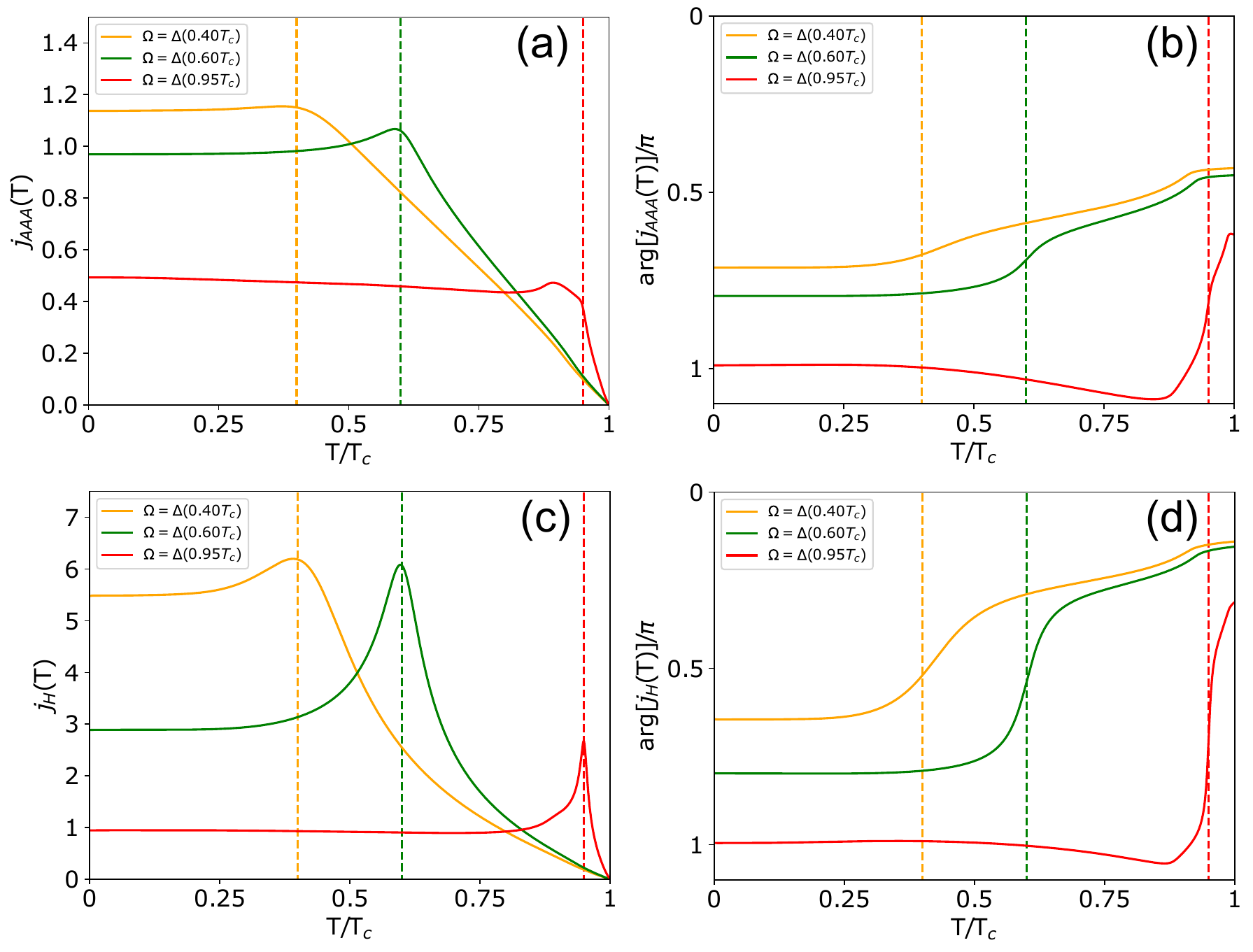}
    \caption{Calculated temperature dependence of the THG currents. In (a) and (b) are the amplitude and phase of $j_{AAA}$ shown. In (c) and (d) is the amplitude and phase of the Higgs contribution $j_H$ shown. We observe that the Higgs contribution dominates the THG current and that both contributions possess a peak and a phase shift at the resonance condition $\Omega=\Delta(T)$ highlighted by the dashed lines.}
    \label{fig:Graph_current_temperature}
\end{figure}

In Fig.\,\ref{fig:Graph_current_temperature} we plot the temperature dependence of the third harmonic current contribution normalized to $j_0 = \sigma D Q^3e^{-1}$. For this, we also separate the amplitude (Higgs) mode contribution $j_{\mathrm{H}}$ and the contribution of the direct action of the electric field $j_{\mathrm{AAA}}$ by separating $\delta\hat{g}=\delta\hat{g}_{Q}+\delta\hat{g}_{\Delta}$ in the Eqs.\,(\ref{CrT4}-\ref{CrT4b}). $\delta\hat{g}_\Delta \propto \delta\hat{\Delta}$ describes the correction to the current arising from the correction to $\Delta$, while $\delta\hat{g}_Q$ describes the direct coupling of the light to the condensate. This separation was already introduced earlier, when we defined $\delta\hat{g}$ in Eqs.\,(\ref{8}) and (\ref{10}).

As discussed in the introduction, one of the most interesting question is whether the amplitude (Higgs) mode contribution dominates over the contribution due to the direct action of ac electric field in the diffusive superconductors. Indeed, we find that the amplitude mode contribution dominates over the direct action of the electric field and this is especially prominent if one looks at the phase shift of the argument of the THG current, consistent with previous calculations by various groups.

\section{Reflection and Transmission Coefficients of ac Field.}

We consider an electromagnetic wave incident on a superconducting film with
thickness $2d$. The reflection (transmission) coefficient is determined from
Maxwell's equations

\begin{eqnarray}
\mathbf{\nabla \times H} &=&\frac{4\pi }{c}\mathbf{j}+\frac{\varepsilon }{c}%
\partial _{t}\mathbf{E}\text{,}  \label{R1} \\
\mathbf{\nabla \times E} &=&-\frac{1}{c}\partial _{t}\mathbf{H}  \label{R2}
\end{eqnarray}%
For a wave of a form $E\infty \exp (i\omega t-ikz)$, these equations can be written as

\begin{eqnarray}
\mathbf{k\times H} &=&\frac{1}{c}(4\pi i\sigma _{\omega }-\varepsilon \omega
)\mathbf{E}\text{,}  \label{R3} \\
\mathbf{k\times E} &=&\frac{\omega }{c}\mathbf{H}\text{,}  \label{R3a}
\end{eqnarray}%

or

\begin{equation}
\mathbf{k\times (k\times E})=\frac{1}{c^{2}}(4\pi i\sigma _{\omega }\omega
-\varepsilon \omega ^{2})\mathbf{E}\text{,}  \label{R3b}
\end{equation}

The dispersion relation $\omega (k)$ follows directly from Eq.\,(\ref{R3b})

\begin{eqnarray}
(ck_{S})^{2} &=&-4\pi i\sigma (\omega )\omega +\varepsilon\omega ^{2}\text{, S film}
\label{R4} \\
(ck_{V})^{2} &=&\varepsilon \omega ^{2}\text{, vacuum}  \label{R4a}
\end{eqnarray}%
Writing down the Maxwell equations for components of $E$ and $H$ we find

\begin{eqnarray}
\partial _{z}H &=&\frac{1}{c}(4\pi \sigma (\omega )+i\omega )E\equiv
\epsilon _{\omega }E\text{.}  \label{R5} \\
\partial _{z}E &\mathbf{=}&\frac{i\omega }{c}H  \label{R5a}
\end{eqnarray}

\begin{widetext}
    
Thus, we obtain for the solution $E(z,t)\equiv E(z)\exp (i\omega t)$ and $%
H(z,t)\equiv H(z)\exp (i\omega t)$ with
\begin{equation}
E(z)={\Big \{}%
\begin{array}{l}
E_{in}\exp (-ik_{V}(z+d))+E_{r}\exp (ik_{V}(z+d))\text{, }z<-d \\
E_{tr}\exp (-ik_{V}(z-d))\text{, } \quad \quad \quad \quad \quad \quad \quad \quad \quad \,\,\,\, z>d%
\end{array}%
{\Big \}}.  \label{R6}
\end{equation}
 
and

\begin{equation}
H(z)=\sqrt{\epsilon _{0}}{\Big \{}%
\begin{array}{l}
-E_{in}\exp (-ik_{V}(z+d))+E_{r}\exp (ik_{V}(z+d))\text{, }z<-d \\
-E_{tr}\exp (-ik_{V}(z-d))\text{, }\quad \quad \quad \quad \quad \quad \quad \quad \quad \,\,\,\, z>d%
\end{array}%
{\Big \}}.  \label{R7}
\end{equation}%

Inside the S film we have

\begin{eqnarray}
E(z) &=&C\cosh (ik_{S}z)+S\sinh (ik_{S}z)\text{, }|z|<d  \label{R8} \\
H(t,z) &=&\sqrt{\epsilon _{\Omega }}[C\sinh (ik_{S}z)+S\cosh (ik_{S}z)]\text{%
, }|z|<d  \label{R8a}
\end{eqnarray}%

\end{widetext}

The matching conditions $[E]_{\pm d}=0$ and $[H]_{\pm d}=0$ yield
\begin{eqnarray}
E_{r}+E_{tr} &=&E_{in}\frac{\sqrt{\epsilon _{0}}-\sqrt{\epsilon _{\omega }}%
\tanh (i\theta _{S})}{\sqrt{\epsilon _{0}}+\sqrt{\epsilon _{\omega }}\tanh
(i\theta _{S})}\text{,}  \label{R9} \\
E_{r}-E_{tr} &=&-E_{in}\frac{\sqrt{\epsilon _{\omega }}-\sqrt{\epsilon _{0}}%
\tanh (i\theta _{S})}{\sqrt{\epsilon _{\omega }}+\sqrt{\epsilon _{0}}\tanh
(i\theta _{S})}  \label{R9a}
\end{eqnarray}%

where $\theta _{S}=k_{S}d$, $k_{S}c=\omega \sqrt{\epsilon _{\omega }}$ and $%
\epsilon _{\omega }=4\pi \sigma (\omega )/i\omega +\epsilon _{S0}$.

As a result we obtain the reflected and transmitted waves

\begin{eqnarray}
E_{r} &=&E_{in}\frac{\tanh (i\theta _{S})(\epsilon _{0}-\epsilon _{\omega })%
}{\mathcal{D}}\text{,}  \label{R10} \\
E_{tr} &=&E_{in}\sqrt{\epsilon _{\omega }\epsilon _{0}}\frac{1}{%
\cosh ^{2}(i\theta _{S})\mathcal{D}}  \label{R10a}
\end{eqnarray}%
where $\mathcal{D}=[\sqrt{\epsilon _{\omega }}+\sqrt{\epsilon _{0}}\tanh
(i\theta _{S})][\sqrt{\epsilon _{0}}+\sqrt{\epsilon _{\omega }}\tanh
(i\theta _{S})]$.

In the limit of a thick S film ($|\theta _{S}|\gg 1$), we obtain $\mathcal{D}%
=[\sqrt{\epsilon _{0}}+\sqrt{\epsilon _{\omega }}]^{2}$ and

\begin{eqnarray}
E_{r} &=&E_{in}\frac{\sqrt{\epsilon _{0}}-\sqrt{\epsilon _{\omega }}}{\sqrt{%
\epsilon _{0}}+\sqrt{\epsilon _{\omega }}}\text{,}  \label{R11} \\
E_{tr} &=&E_{in}\frac{\sqrt{\epsilon_0 \epsilon_\omega}}{%
\cosh ^{2}(i\theta _{S})[\sqrt{\epsilon _{0}}+\sqrt{\epsilon _{\omega }}]^{2}%
}  \label{R11a}
\end{eqnarray}%
As expected, Eq.\,(\ref{R11a}) shows that the amplitude of the transmitted wave is
exponentially small: $E_{tr}\infty E_{in}\exp (-2d/d_{skin})$, where $%
d_{skin}=c/\sqrt{4\pi i\sigma (\omega )\omega }$ is the skin depth.

\begin{figure} [htbp]
    \centering
    \includegraphics[width=0.49 \textwidth]{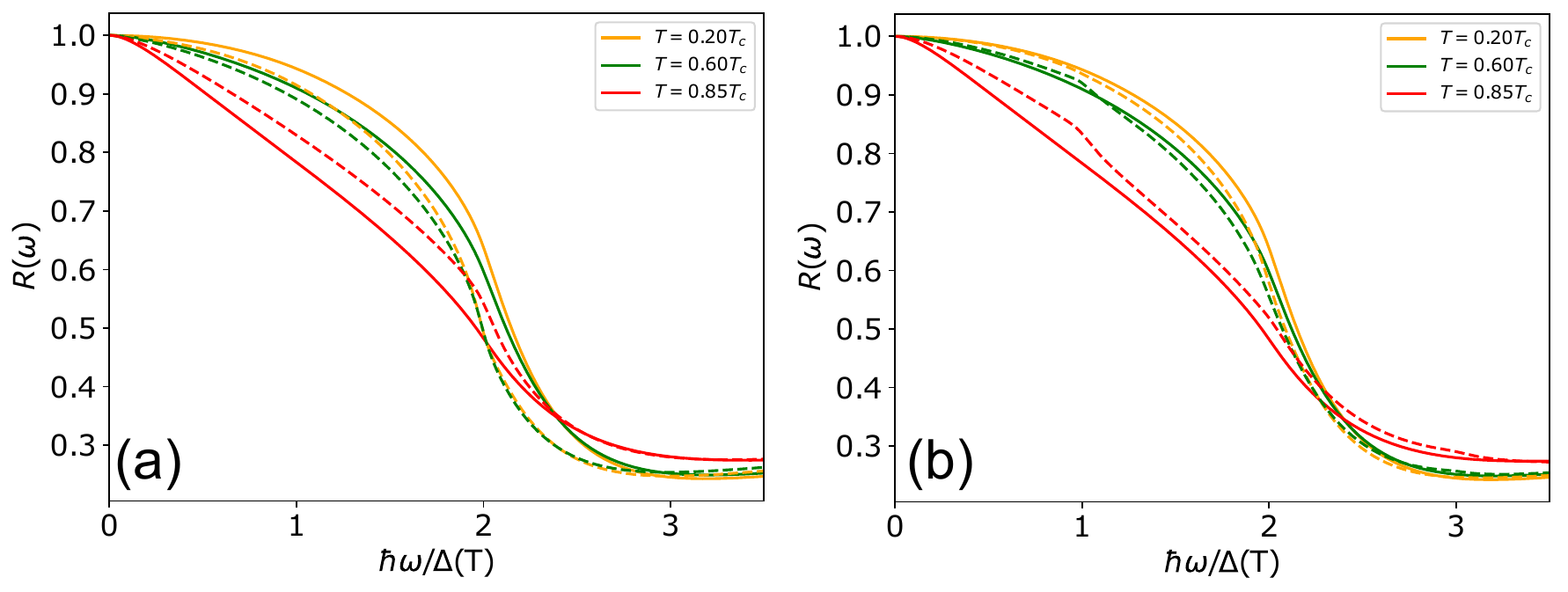}
    \caption{The reflectivity $R(\omega)=\frac{|E_r|^2}{|E_{in}|^2}$ via equation\,(\ref{R10}). The solid lines represent the case of the equilibrium, while the dashed lines are obtained by taking the correction $\delta\sigma(\omega, \Omega)$ into account when determining $\epsilon_\omega$. The pump-frequency is fixed at $\Omega=2\Delta(T)$ for (a) and $\Omega=\Delta(T)$ for (b). Further, $DQ^2=0.005\Delta(T)$, $\gamma=0.02\Delta_0$, $\sigma_0=2\times10^{4}\,\mathrm{cm}^{-1}\mathrm{\Omega}^{-1}$ and a film-thickness of $d=12\,\mathrm{nm}$. For a pump of $\Omega=2\Delta(T)$, we observe that the high-temperature reflectivity curve is getting enhanced under irradiation, while this is not the case for the lower-temperature curves. Lastly, we observe that for a lower pump-frequency $\Omega=\Delta(T)$ we see a small area of reflectivity enhancement for the lower-temperature curve of $T=0.6T_c$, which is due to the redistribution of particles inside the gap and not due to the Eliashberg effect.}
    \label{fig:Graph_Reflectivity_Eliashberg}
\end{figure}

For a thin S film ($|\theta_s|\ll1$) we plot the reflectivity $R(\omega)=\frac{|E_r|^2}{|E_{in}|^2}$ as a function of normalized frequency for various temperatures in Fig.\ref{fig:Graph_Reflectivity_Eliashberg}. Observe clear signatures of the Eliashberg effect for temperatures close to the superconducting transition temperature for both $\Omega=\Delta(T)$ and $\Omega=2\Delta(T)$. 

Next we look into the effects of parametric amplification. For that we assumed that the EM-field inside the thin S-film should be either uniform or slowly varying function. Thus, we take the average of the electric-field inside the film:
\begin{equation}\label{E_thin_film}
\begin{split}
    \langle E_\omega^S\rangle = \frac{E_{in}\tan{(\theta_s)}}{2\theta_s}\left(1+\frac{\sqrt{\epsilon_V}-\sqrt{\epsilon_\omega}\tanh{(i\theta_s)}}{\sqrt{\epsilon_V}+\sqrt{\epsilon_\omega}\tanh{(i\theta_s)}}\right)
\end{split}
\end{equation}
Most importantly, we can now link the incoming electric-field $E_{in}(\omega)$ with the outgoing field $E_{r}(2\Omega-\omega)=E_{tr}(2\Omega-\omega)=E_{2\Omega-\omega}$ via the THG current
\begin{equation}
    j(2\Omega-\omega)=\frac{\Lambda_{ac}(2\Omega-\omega)}{i\omega}\langle E^S_\omega\rangle = \sigma^{(1)}(2\Omega-\omega)E_{2\Omega-\omega}.
\end{equation}
Using the definition from Eq.\,(\ref{E_thin_film}) for the field $E_\omega$ inside of the S film we define the downconversion intensity
\begin{equation}\label{R12a}
\begin{split}
    R_{12}(2\Omega-\omega)=&\left|\frac{\Lambda_{ac}(2\Omega-\omega)}{i\omega\sigma^{(1)}(2\Omega-\omega)}\right|^2\\&\times\left|\frac{\tan{(\theta_s)}}{2\theta_s}\left(1+\frac{\sqrt{\epsilon_V}-\sqrt{\epsilon_\omega}\tanh{(i\theta_s)}}{\sqrt{\epsilon_V}+\sqrt{\epsilon_\omega}\tanh{(i\theta_s)}}\right)\right|^2
\end{split}
\end{equation}
which is similar to Ref.\,\cite{Buzzi2021PRX} with the important difference that all quantities, entering its definition are now determined fully microscopically. In particular,
the function $\Lambda_{ac}$ is directly given by the THG current $j(2\Omega-\omega)$ and $\sigma^{(1)}$ is the linear-response of the complex conductivity. 

\begin{figure} [htbp]
    \centering
    \includegraphics[width=0.49 \textwidth]{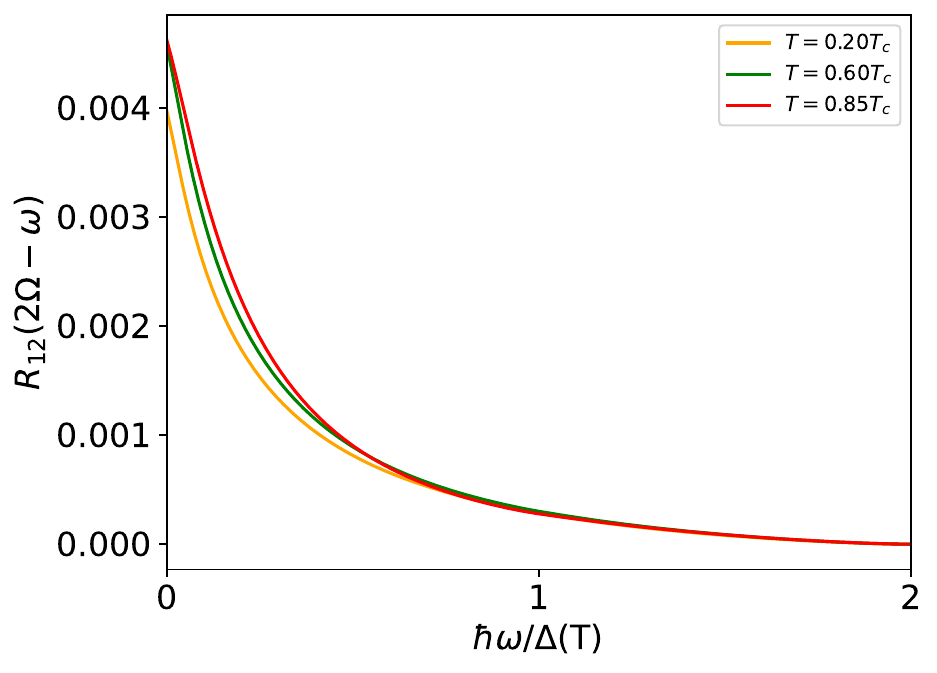}
    \caption{The downconversion intensity $R_{12}(2\Omega-\omega)$  i.\,e.\,$\frac{|\vec{E}_{2\Omega-\omega}|^2}{|\vec{E}_\omega|^2}$, as a function of the probe-frequency $\omega$. The pump-frequency is fixed at $\Omega=\Delta(T)$ and $DQ^2=0.005\Delta(T)$. Further, $\gamma=0.02\Delta_0$, $\sigma_0=2\times10^{4}\,\mathrm{cm}^{-1}\mathrm{\Omega}^{-1}$ and a film-thickness of $d=12\,\mathrm{nm}$. The highest intensity is achieved for $\omega\rightarrow0$, while it vanishes at the points $\omega=2\Omega$. For frequencies of $\Delta(T)\leq\omega\leq2\Delta(T)$ the curves visually overlay.}
    \label{fig:Graph_Reflectivity_downsized}
\end{figure}

The down-conversion intensity $R_{12}(2\Omega-\omega)$ for $\Omega=\Delta(T)$ is shown in Fig.\,\ref{fig:Graph_Reflectivity_downsized} as function of the probe-frequency $\omega$. It is noted, that in contrast to $R(\omega)$ we do not include the conductivity correction $\delta\sigma(\omega, \Omega)$ in the definition of $\epsilon_\omega$ in Eq.\,(\ref{R12a}).  Similar to the phenomenological analysis, presented in Ref.\,\cite{Buzzi2021PRX}, we observe that the down-conversion intensity decreases rapidly in the region of $\omega<2\Delta(T)$ and that it vanishes at $\omega=2\Delta(T)$. Note, that in Ref.\,\cite{Buzzi2021PRX} this quantity was related to the amplitude of the emitted idler mode, normalized by the amplitude of the incident signal beam and connected to the parametric amplification of superconductivity due to the Higgs mode. 

In our case, we can indeed confirm that the behavior of the down-conversion intensity is expected due to the third harmonic generation currents and excitation of the Higgs mode and the direct action of the electric field on the charge carriers in the superconducting state. 

\section{Conclusions}

To conclude, we have developed a theory of non-linear effects arising in diffusive superconductors under the action of an $ac$ electromagnetic
fields $E(t)$. These effects are described in terms of quasiclassical matrix
Green's functions $\hat{g}$ which consists of the retarded (advanced), $\hat{g}^{R(A)}$, and Keldysh Green's functions \cite{KopninBook01}. We use a
method of the representation of the Keldysh function $\hat{g}^{K}$ as the
sum of a regular, $\hat{g}^{reg}$, and "anomalous", $\hat{g}^{an}$, parts \cite{GorkovEliashberg68}. This trick allows one to avoid the method of
analytical continuation \cite{KopninBook01}. Using this representation, we
derive general expressions for $\hat{g}^{R(A)}$ and $\hat{g}^{an}$. On the
basis of this formalism, we obtain the variation of the order parameter $\delta\Delta$
caused by the ac field up to the second order in the magnitude of $E(t)$. In particular, we analyze the zero Fourier harmonic 
of $\delta\Delta_{0}$ (the Eliashberg effect \cite{Eliashberg70}) and the second
Fourier harmonic $\delta\Delta_{2}$.  
Furthermore, we calculated the ac currents $I(\Omega)$ induced by external electromagnetic fields. Analyzing the third harmonic generating current we indeed confirm that in the diffusive superconductors it is mostly dominated by the amplitude ("Higgs") mode and not by the direction action of the ac electric field. This is contrast to the clean case.

Finally, we also analyze microscopically the down-conversion intensity, $R_{12}$, of the thin superconducting sample, which characteristic behavior was argued to be related to the parametric amplification of superconductivity. Although studying parametric amplification goes beyond the present theoretical analysis we indeed see that a very similar behavior of $R_{12}$ is expected due to the coupling to the amplitude mode and the direct action of the electric field in the third harmonic generation currents. 
We obtained also a strong enhancement of a photo-excited ac conductance which occurs at low frequencies and low temperatures. This issue deserves separate consideration.

Note, that our formalism for calculation of the non-linear currents can be used to analyse the data on
transient transport in pumped conventional and unconventional superconductors.

\section{Acknowledgements}
We are thankful to Maxim Dzero for careful reading of the manuscript.
The work is supported by the Mercator Research Center Ruhr (MERCUR) project "Composite collective excitations in correlated quantum materials" of the University Alliance Ruhr.

\section{Appendix}
\subsection{Anomalous Green's Function}\label{App:AnoGreen}

The Keldysh component of Eq.\,(\ref{6}) can be written as
\begin{equation}
\hat{N}_{\epsilon }\delta \hat{g}^{K}-\delta \hat{g}^{K}\hat{N}_{\epsilon
^{\prime }}=-\mathcal{\hat{R}}^{K}(\epsilon ,\epsilon ^{\prime })
\label{Ap1}
\end{equation}%
where $\hat{N}_{\epsilon }=\epsilon \hat{\tau}_{3}+\Delta i\hat{\tau}_{2}$, $%
\mathcal{\hat{R}}^{K}(\epsilon ,\epsilon ^{\prime })=\{[\delta \hat{\Delta},%
\hat{g}]+iD[Q\hat{\tau}_{3}\hat{g}Q\hat{\tau}_{3},\hat{g}]\}^{K}$. The
normalisation condition for the Keldysh component is

\begin{equation}
\delta \{\hat{g}^{R}\hat{g}^{K}+\hat{g}^{K}\hat{g}^{A}\}_{\epsilon ,\epsilon
^{\prime }}=0  \label{Ap2a}
\end{equation}%
or

\begin{equation}
\delta \hat{g}^{R}\hat{g}_{\epsilon ^{\prime }}^{K}+\hat{g}_{\epsilon
}^{R}\delta \hat{g}^{K}+\delta \hat{g}^{K}\hat{g}_{\epsilon ^{\prime }}^{A}+%
\hat{g}_{\epsilon }^{K}\delta \hat{g}^{A}=0  \label{Ap2b}
\end{equation}
Here $\hat{g}_{\epsilon }^{K}=(\hat{g}_{\epsilon }^{R}-\hat{g}_{\epsilon
}^{A})t_{\epsilon }$ is the Keldysh Green's function in equilibrium. The variation of the Keldysh function $\delta \hat{g}^{K}$ is represented as a
sum of a regular part and the anomalous part $\delta \hat{g}^{an}$ (see
Eq.\,(\ref{9}))

\begin{equation}
\delta \hat{g}^{K}=\delta \hat{g}^{R}t_{\epsilon ^{\prime }}-t_{\epsilon
}\delta \hat{g}^{A}+\delta \hat{g}^{an}  \label{Ap3}
\end{equation}%
Taking into account Eq.\,(\ref{Ap3}), we can write Eq.\,(\ref{Ap2b}) as
\begin{eqnarray}
&&0=(\delta \hat{g}^{R}\hat{g}_{\epsilon ^{\prime }}^{R}+\hat{g}_{\epsilon}^{R}\delta \hat{g}^{R}%
)t_{\epsilon ^{\prime }}-t_{\epsilon
}(\delta \hat{g}^{A}\hat{g}_{\epsilon ^{\prime }}^{A}+\hat{g}_{\epsilon}^{A}\delta \hat{g}^{A}%
)\nonumber \\
&&+\hat{g}_{\epsilon }^{R}\delta \hat{g}^{an}+\delta
\hat{g}^{an}\hat{g}_{\epsilon ^{\prime } }^{A} \label{Ap4}  
\end{eqnarray}%

The first two terms are equal to zero due to the variation of the
normalization conditions for the matrices $\hat{g}^{R(A)}$

\begin{equation}
(\delta \hat{g}\hat{g}_{\epsilon ^{\prime }}+\hat{g}_{\epsilon }\delta
\hat{g})^{R(A)}=0  \label{Ap5}
\end{equation}%
Thus, we come to the equation for $\delta \hat{g}^{an}$

\begin{equation}
\hat{g}_{\epsilon }^{R}\delta \hat{g}^{an}+\delta \hat{g}^{an}\hat{g}%
_{\epsilon ^{\prime }}^{A}=0.  \label{Ap6}
\end{equation}

Then, we multiply Eq.\,(\ref{6}) for $\delta \hat{g}^{R(A)}$ by $t_{\epsilon
^{\prime }}$ ($t_{\epsilon }$) and subtract (add) from (to) Eq.\,(\ref{Ap1}).
Taking into account the definition of the matrix $\hat{N}_{\epsilon
}^{R(A)}=(\zeta _{\epsilon }\hat{g}_{\epsilon })^{R(A)}$, we obtain

\begin{eqnarray}
&&(\zeta _{\epsilon }\hat{g})^{R}\delta \hat{g}^{an}-\delta \hat{g}%
^{an}(\zeta _{\epsilon ^{\prime }}\hat{g}_{\epsilon ^{\prime }})^{A}
\nonumber \\
&=&-\{\mathcal{\hat{R}}^{K}(\epsilon ,\epsilon ^{\prime })-\mathcal{\hat{R}}%
^{R}(\epsilon ,\epsilon ^{\prime })t_{\epsilon ^{\prime }}+t_{\epsilon }%
\mathcal{\hat{R}}^{A}(\epsilon ,\epsilon ^{\prime })\}\label{Ap7} 
\end{eqnarray}%

The matrix $\mathcal{\hat{R}}^{K}(\epsilon ,\epsilon ^{\prime })$ contains
only the Green' s function in equlibrium when $\hat{g}_{\epsilon }^{K}=(%
\hat{g}_{\epsilon }^{R}-\hat{g}_{\epsilon }^{A})t_{\epsilon }$. Using the
normalization condition, Eq.\,(\ref{Ap6}), we find for the anomalous function $%
\delta \hat{g}^{an}$
\begin{equation}
\delta \hat{g}^{an}=-\hat{g}_{\epsilon }^{R}\frac{\{\mathcal{\hat{R}}%
^{K}(\epsilon ,\epsilon ^{\prime })-\mathcal{\hat{R}}^{R}(\epsilon ,\epsilon
^{\prime })t_{\epsilon ^{\prime }}+t_{\epsilon }\mathcal{\hat{R}}%
^{A}(\epsilon ,\epsilon ^{\prime })\}}{\zeta _{\epsilon }^{R}+\zeta
_{\epsilon ^{\prime }}^{A}}  \label{Ap8}
\end{equation}

Substituting the known Green's functions into Eq.\,(\ref{Ap8}) we come to Eq.\,(%
\ref{10}-\ref{10a}).
\subsection{Useful functions}
We list the definition of useful functions to avoid cluttering in the definition of $\delta\hat{g}_{\epsilon,\epsilon'}$
\begin{eqnarray}
 N_{3}^{R(A)} &=&(A_{\epsilon }^{(-)}g_{\epsilon +\Omega }-B_{\epsilon}f_{\epsilon +\Omega })^{R(A)}\text{,}  \label{CrE7a} \\
N_{2}^{R(A)} &=&(A_{\epsilon }^{(+)}f_{\epsilon +\Omega }+B_{\epsilon
}g_{\epsilon +\Omega })^{R(A)}\text{.}  \label{CrE7b}
\end{eqnarray}

The functions $N_{2,3}^{an}$, $G_{\tilde{\epsilon}}^{(-)}$ and $F_{\tilde{ \epsilon}}^{(-)}$ are defined as follows

 \begin{eqnarray}
N_{3}^{an} &=&G_{\tilde{\epsilon}}^{(-)}A_{\epsilon }^{(-)}+F_{\tilde{%
\epsilon}}^{(-)}B  \label{CrE8a} \\
N_{2}^{an} &=&F_{\tilde{\epsilon}}^{(-)}A_{\epsilon }^{(+)}+G_{\tilde{%
\epsilon}}^{(-)}B\text{,}  \label{CrE8B} \\
 G_{\epsilon +\Omega }^{(-)} &=&g_{\epsilon +\Omega }^{R}-g_{\epsilon +\Omega
}^{A},F_{\epsilon +\Omega }^{(-)}=f_{\epsilon +\Omega }^{R}-f_{\epsilon +\Omega }^{A},  \label{CrE8c}
\end{eqnarray}%

The functions $A$, $B$ and $C$ are

\begin{eqnarray}
(A_{\epsilon }^{\pm })^{R} &=&(1\pm (g_{\epsilon }^{2}+f_{\epsilon}^{2}))^{R}\text{,}  \label{CrE10} \\
 B_{\epsilon }^{R} &=&2(g_{\epsilon }f_{\epsilon })^{R}\text{,}
 \label{CrE10a} \\
C_{\epsilon }^{R} &=&2\zeta _{\epsilon }^{R}  \label{CrE10b}
 \end{eqnarray}%
Anomalous functions $A_{\epsilon }^{an},B_{\epsilon }^{an},C_{\epsilon}^{an} $ are

\begin{eqnarray}
(A_{\epsilon }^{\pm })^{an} &=&1\pm (g_{\epsilon }^{R}g_{\epsilon}^{A}+f_{\epsilon }^{R}f_{\epsilon }^{A})\text{,}  \label{CrE11} \\
B_{\epsilon }^{an} &=&g_{\epsilon }^{R}f_{\epsilon }^{A}+g_{\epsilon}^{A}f_{\epsilon }^{R}\text{,}  \label{CrE11a} \\
C_{\epsilon }^{an} &=&\zeta _{\epsilon }^{R}+\zeta _{\epsilon }^{A}\text{.}
\label{CrE11b}
 \end{eqnarray}

Here we provide further expressions, which are useful in obtaining the final expressions for the Eliashberg effect and the third harmonic generation currents in the main text.

\bigskip For the $\delta \Delta (t)$ averaged in time we have
\begin{equation}
\delta \Delta =-\lambda i\int \frac{d\epsilon }{2\pi }\text{Tr}\hat{\tau}%
_{2}\{\delta \hat{g}^{reg}+\hat{g}^{an}\}|_{\epsilon _{-}=0}  \label{AE1}
\end{equation}%
Taking into account the identity
\begin{equation}
\delta \Delta =\lambda \delta \Delta \int \frac{d\epsilon }{2\pi }\left[\frac{1}{%
\zeta _{\epsilon }^{R}}-\frac{1}{\zeta _{\epsilon }^{A}}\right]t_{\epsilon }
\label{AE2}
\end{equation}%
and subtracting Eqs.\,(\ref{AE1},\ref{AE2}), we obtain
\begin{equation}
0=\sum_{\nu ,\mu }\int \frac{d\epsilon }{2\pi }\left[\text{Tr}\{i\hat{\tau}%
_{2}(\delta \hat{g}^{reg}+\hat{g}^{an})\}+\left(\frac{1}{\zeta _{\epsilon }^{R}}-%
\frac{1}{\zeta _{\epsilon }^{A}}\right)t_{\epsilon }\delta \Delta \right]  \label{AE3}
\end{equation}%
where $\zeta _{\epsilon }^{R(A)}=\sqrt{(\epsilon \pm i\gamma )^{2}+\Delta
^{2}}$. This equation can be written in the form
\begin{equation}
4T\delta \Delta _{0}\Delta ^{2}\sum_{n}\frac{1}{\zeta _{n}^{3}}%
=-i(DQ^{2})\int \frac{d\epsilon }{2\pi }\frac{\delta f_{Q}^{reg}+\delta
f_{Q}^{an}}{2\zeta _{\epsilon }}\text{.}  \label{AE3a}
\end{equation}%
where $\zeta _{n}=\sqrt{\epsilon _{n}^{2}+\Delta ^{2}}$, $\epsilon _{n}=\pi
T(2n+1)$. We used Eq.\,(\ref{E2a}) with $f_{\Delta }=1+g_{n}^{2}+f_{n}^{2}$, $%
g_{n}^{2}=(\epsilon _{n}/\zeta _{n})^{2}=1-f_{n}^{2}$.

The regular part at the right is

\begin{eqnarray}
\int \frac{d\epsilon }{2\pi }\frac{\delta f_{Q}^{reg}}{2\zeta _{\epsilon }}
&=&\int \frac{d\epsilon }{2\pi }\left[\left(\frac{A_{+}f_{\epsilon +\Omega
}+Bg_{\epsilon +\Omega }}{2\zeta _{\epsilon }}\right)^{R}-(..)^{A}\right]t_{\epsilon }
\label{AE4} \nonumber \\
&=&\int \frac{d\epsilon }{2\pi }\left[\left(\frac{g_{\epsilon }(g_{\epsilon
}f_{\epsilon +\Omega }+f_{\epsilon }g_{\epsilon +\Omega })}{\zeta _{\epsilon
}}\right)^{R}-(..)^{A}\right]t_{\epsilon}  \nonumber\\
& = & \Delta \int \frac{d\epsilon }{2\pi }\left[ \left( \frac{\epsilon (2\epsilon +\Omega )}{\zeta _{\epsilon }^{3}\zeta _{\epsilon +\Omega }}\right)^{R}-(..)^{A}\right] t_{\epsilon} \text{.}
\end{eqnarray}%
\newline

We express the integral in terms of Matsubara frequencies

\begin{eqnarray}
\int \frac{d\epsilon }{2\pi }\frac{\delta f_{Q}^{reg}}{2\zeta _{\epsilon }}
&=&\Delta \int \frac{d\epsilon }{2\pi }\left[\left(\frac{\epsilon (2\epsilon +\Omega )
}{\zeta _{\epsilon }^{3}\zeta _{\epsilon +\Omega }}\right)^{R}-(..)^{A}\right]
\label{AE4a} \\
&=&-4iT2\Delta \text{Re}\sum_{n}\frac{\omega (2\omega +i\Omega )}{\zeta
_{\omega }^{3}\zeta _{\omega +i\Omega }}\text{.}
\end{eqnarray}

We took into account the term with $-\Omega $. Therefore, Eq.\,(\ref{E3a})
yields

\begin{equation}
\begin{split}
4T\delta \Delta _{0}\Delta ^{2}\sum_{n}\frac{1}{\zeta _{n}^{3}}=-&8T\Delta DQ^2
\text{Re}\sum_{n}\frac{\omega (2\omega +i\Omega )}{\zeta _{\omega }^{3}\zeta
_{\omega +i\Omega }}\\ &-iDQ^{2}\int \frac{d\epsilon }{2\pi }\frac{\delta
f_{Q}^{an}}{2\zeta _{\epsilon }}\text{.}  \label{AE5}
\end{split}
\end{equation}

\subsection{Third Harmonic}

The response to an external ac field with the frequency $\Omega _{in}$ is

 \begin{equation}
I(t)\text{ }=\sigma \text{ }\exp (i(2\Omega +\Omega _{in})t)Q_{in}\int \frac{%
d\epsilon }{2\pi }J(\epsilon )\text{,}
\end{equation}%
where
\begin{eqnarray}
 J &=&\int \frac{d\epsilon }{2\pi} \left[ \delta \hat{g} \langle \hat{g}_{\epsilon +2\Omega +\omega} \rangle + \langle \hat{g}_{\epsilon -\omega}
\rangle \delta \hat{g}_{0}^{K} \right] \nonumber \\
 &=&\int \frac{d\epsilon }{2\pi }\{\delta \hat{g}^{R}[\langle \hat{g}%
 _{\epsilon +2\Omega +\omega }^{R}-\hat{g}_{\epsilon +2\Omega +\omega
 }^{A}\rangle t_{\epsilon +2\Omega +\omega }  \nonumber \\
 &&+\left[\delta \hat{g}^{R}t_{\epsilon ^{\prime }}-t_{\epsilon }\delta \hat{g}%
^{A}\right] \langle \hat{g}_{\epsilon +2\Omega +\omega }^{A}\rangle  \nonumber  \\
 &&\langle \hat{g}_{\epsilon -\omega }^{R}\rangle \lbrack \delta \hat{g}%
 ^{R}t_{\epsilon ^{\prime }}-t_{\epsilon }\delta \hat{g}^{A}]+\langle \hat{g}%
 _{\epsilon -\omega }^{R}-\hat{g}_{\epsilon -\omega }^{A}\rangle t_{\epsilon
 -\omega }\delta \hat{g}^{A}]  \nonumber \\
 &&+\delta \hat{g}^{an}[\langle \hat{g}_{\epsilon +2\Omega +\omega }^{R}+%
 \hat{g}_{\epsilon -\omega }^{A}\}_{0}\text{.}
 \end{eqnarray}

 \subsection{Coefficients in the Current (EE and Third Harmonic)}

 If $\Omega _{in}=\Omega $, the total current is

 \begin{equation}
 I_{3\Omega }(t)\text{ }=\sigma Q\exp (3i\Omega t)\int \frac{d\epsilon }{2\pi
 }\{\langle \hat{g}_{\epsilon -\Omega }\rangle \cdot \delta \hat{g}+\delta
 \hat{g}\cdot \langle \hat{g}_{\epsilon +3\Omega }\rangle \}_{0}^{K}\text{.}
 \label{Cr3}
 \end{equation}%
 where $\delta \hat{g}=\delta \hat{g}(\epsilon ,\epsilon +2\Omega )$. 
 Here the matrix $\delta \hat{g}_{\mp }^{an}\equiv \delta \hat{g}^{an}(\tilde{%
 \epsilon}_{-},\tilde{\epsilon}_{+})$ is defined as
 \begin{eqnarray}
 \hat{g}_{\Delta }^{an} &=&-r_{\epsilon }\frac{\sinh (2\Omega \beta )}{2}%
 \frac{\delta \Delta _{2\Omega }}{C_{\mp }^{an}}(g_{\Delta }^{an}\hat{\tau}%
 _{3}+f_{\Delta }^{an}i\hat{\tau}_{2})\text{,}  \label{Cr5} \\
 \hat{g}_{Q}^{an} &=&-r_{\epsilon }\frac{\sinh (2\Omega \beta )}{2}\frac{%
 iDQ^{2}}{C_{\mp }^{an}}(g_{Q}^{an}\hat{\tau}_{3}+f_{Q}^{an}i\hat{\tau}_{2})
 \label{Cr5A}
 \end{eqnarray}%
 Thus, for $J^{an}$ we find

 \begin{equation}
 J^{an}=\{(g_{\tilde{\epsilon}-2\Omega }^{R}+\hat{g}_{\tilde{\epsilon}%
 +2\Omega }^{A})(g_{\Delta }^{an}+g_{Q}^{an})+(f_{\tilde{\epsilon}-2\Omega
 }^{R}+f_{\tilde{\epsilon}+2\Omega }^{A})(f_{\Delta }^{an}+f_{Q}^{an}))
 \label{Cr5B}
 \end{equation}%
 The coefficients $g_{\Delta ,Q}^{an}$ and $f_{\Delta ,Q}^{an}$ are defined as

 \begin{eqnarray}
 g_{\Delta }^{an} &=&B\text{, }f_{\Delta }^{an}=A_{+}\text{,}  \label{Cr6a} \\
 g_{Q}^{an} &=&[G_{+}A_{-}-F_{+}B-t_{\epsilon }t_{\Omega }(G_{-}A_{-}-F_{-}B)]%
 \text{,}  \label{Cr6b} \\
f_{Q}^{an} &=&[F_{+}A_{+}+G_{+}B-t_{\epsilon }t_{\Omega
 }(F_{-}A_{+}+G_{-}B)])\text{.}  \label{Cr6c}
 \end{eqnarray}%
 The coefficients $A,B,C$ are equal to
 \begin{eqnarray}
 A_{\pm } &=&1\pm (g_{-}^{R}g_{+}^{A}+f_{-}^{R}f_{+}^{A})\text{,}  \label{Cr7}
 \\
 B &=&g_{-}^{R}f_{+}^{A}+f_{-}^{R}g_{+}^{A}\text{,}  \label{Cr7a}
 \end{eqnarray}%
The functions $G_{\pm },F_{\pm }$ are defined as

 \begin{equation}
G_{\pm }=g_{\epsilon }^{R}\pm g_{\epsilon }^{A}\text{, }F_{\pm }=f_{\epsilon
 }^{R}\pm f_{\epsilon }^{A}  \label{Cr8}
 \end{equation}

\bibliography{references}

\end{document}